\documentclass[preprint2]{emulateapj}
\usepackage{natbib,graphicx,amsmath,longtable,amssymb,CJK} 

\def\OVI{\ion{O}{6}~}   
\def\HI{\ion{H}{1}~}   
   
\def\SiIII{\ion{Si}{3}~}   
\def\SiII{\ion{Si}{2}~}

\def\CII{\ion{C}{2}~}   
\def\OI{\ion{O}{1}~}   
\def\msun{\mbox{$\rm M_\odot$}}
\def\lya{Ly$\alpha$~}
\def\nhi{\mbox{$N_{\rm HI}$}}
\def\h100{\mbox{$h^{\rm -1}$}}
\def\kms{\mbox{$\rm ~km~s^{-1}$}}
\def\cm-2{\mbox{$\rm ~cm^{-2}$}}
\def\mA{\mbox{$\rm m\AA$}}
\def\Wprime{W$^{\prime}$~}

\shorttitle{Gas in the Virgo Cluster }
\shortauthors{Joo Heon Yoon}
\slugcomment{5/17/2012}

\begin{document}

\title{Warm Gas in the Virgo Cluster: I. Distribution of \lya Absorbers}

\author{Joo Heon Yoon\altaffilmark{1}$^{\star}$, Mary E. Putman\altaffilmark{1}, Christopher Thom\altaffilmark{2}, Hsiao-Wen Chen\altaffilmark{3}, Greg L. Bryan\altaffilmark{1}}

\altaffiltext{1}{Department of Astronomy, Columbia University, New York 10027, USA}
\email{$^{\star}$jhyoon@astro.columbia.edu}
\altaffiltext{2}{Space Telescope Science Institute, 3700 San Martin Dr, Baltimore, MD 21211, USA}
\altaffiltext{3}{Department of Astronomy \& Astrophysics and Kavli Institute for Cosmological Physics, University of Chicago, Chicago, IL 60637, USA}

\begin{abstract}
The first systematic study of the warm gas ($T=10^{4-5} \rm K$) distribution across a galaxy cluster is presented using multiple background QSOs to the Virgo Cluster.  We detect 25 \lya absorbers ($\nhi = 10^{13.1-15.4}\cm-2$) in the Virgo velocity range toward 9 of 12 QSO sightlines observed with Cosmic Origin Spectrograph, with a cluster impact parameter range of $0.36-1.65$ Mpc ($0.23-1.05R_{\rm vir}$).  Including 18 \lya absorbers previously detected by STIS or GHRS toward 7 of 11 background QSOs in and around the Virgo Cluster,
we establish a sample of 43 absorbers toward a total of 23 background probes for studying the incidence of \lya absorbers in and around the Virgo Cluster. With these absorbers, we find: (1) warm gas is predominantly in the outskirts of the cluster and avoids the X-ray detected hot intracluster medium (ICM). Also, \lya absorption strength increases with cluster impact parameter.  (2) \lya absorbing warm gas traces cold \HI emitting gas in the substructures of the Virgo Cluster. (3) Including the absorbers associated with the surrounding substructures, the warm gas covering fraction (100\% for $\nhi > 10^{13.1}\cm-2$) is in agreement with cosmological simulations. We speculate that the observed warm gas is part of large-scale gas flows feeding the cluster both the ICM and galaxies. 
\end{abstract}

\section{Introduction}

Most of the baryons in the universe exist in diffuse gas phases \citep{Dave2010a, Shull2011a}. Simulations predict that the diffuse gas that produces \lya absorbers ($T<10^5 \rm~K$) occupies most of the cosmic volume and accounts for 20\%--40\% of the baryonic mass budget at $z=0$ \citep{Cen1999a, Dave2001a}. Observations of low redshift \lya clouds find 20\%--30\% of the baryons in photoionized warm gas which is consistent with the simulations \citep{Penton2000a,Penton2004a,Lehner2007a,Danforth2008a}. 

In a galaxy cluster, gas can exist in a variety of phases. Studies of the kinematics and thermal state of the ICM provide important clues about clusters, galaxies, and star formation mechanisms in cosmological simulations \citep{Loken2002a,Kravtsov2005a,Roncarelli2006a,Burns2010a}.  Most of the gas falling into a cluster's dark matter potential well is shock heated to the virial temperature \citep{White1978a,Cen1999a,Dave1999a}, and X-ray emission from this hot ICM is commonly observed. The hot gas contributes $\sim80\%$ of the total baryonic content while the contribution of cold neutral gas (observed via 21 cm emission) is less than 1\% in a galaxy cluster \citep{Ettori2004a}. The warm gas ($T=10^{4-5} \rm~K$) remains largely unprobed in a galaxy cluster despite its large mass fraction in the universe, and importance in understanding the multiphase nature of the ICM.
  It is too cold to be probed by X-ray emission, too hot to be traced by \HI emission, and usually too diffuse to be detected in optical emission.  
Observations of \lya absorption lines are one of the most promising methods to study warm gas in the intergalactic medium \citep[IGM;][and references therein]{Rauch1998a}.

Absorption line studies have been used to probe a variety of structures in the universe.
The early work by \citet{Spitzer1956a} suggested the possible existence of a galactic corona surrounding a galaxy which could result in multiple UV absorption lines in a QSO spectrum \citep{Bahcall1969a}.   While some \lya absorbers are thought to be directly linked to halo gas in galaxies \citep{Carilli1992a,Lanzetta1995a,Gorkom1996a,Chen1998a,Chen2001a,Penton2002a}, others seem to be unrelated to any galaxy systems \citep[see the review of][]{Rauch1998a}.  Rather, they seem to be associated with structures in the IGM, such as filaments or sheets \citep{Dinshaw1997a,Impey1999a,Rosenberg2003a,Cote2005a}. More specifically, weak absorbers ($W<100-300~\mA $) do not cluster strongly with known galaxies and the majority are speculated to originate in the cosmic web \citep{Chen2009a,Prochaska2011a}. 
Attempts to map  warm gas in large scale structures between galaxies and filaments (i.e., groups and clusters) are limited. IGM absorbers associated with a galaxy group or cluster toward a single QSO sightline were studied by several authors \citep{Lanzetta1996a,Koekemoer1998a,Tripp1998a,Ortiz-Gil1999a,Miller2002a,Prochaska2006a,Lopez2008a}, but no studies of the warm gas distribution throughout a cluster have been made.

The Virgo Cluster is our closest galaxy cluster \citep[16.5 Mpc,][]{Mei2007a} and is therefore very extended spatially ($\sim 12^{\circ}$ diameter). It provides a unique location to study the flow of gas in relation to both small and large dark matter structures. Observations of this irregular cluster have been obtained at various wavelengths to map the X-ray emitting hot gas \citep[e.g.,][]{Bohringer1994a,Urban2011a}, atomic hydrogen \citep{Gavazzi2005a,Giovanelli2007a, Popping2011a}, dust \citep{Davies2010a}, and stars \citep{Cote2004a,Mihos2005a}.  The Virgo Cluster therefore provides a unique laboratory for studying the relationship between baryons in different phases.

The ability to map warm gas in relation to dark matter structures has increased substantially with access to the Cosmic Origin Spectrograph \citep[COS, ][]{Green2012a} onboard the {\it Hubble Space Telescope (HST).}  It has a higher throughput than previous instruments, allowing us to obtain high signal to noise UV spectra of faint QSOs and increasing the surface density of background probes available for absorption line spectroscopy. 
In this paper, the warm gas distribution in and around a single galaxy cluster is studied with multiple sightlines systematically for the first time.  The data enable us to probe the spatial distribution of the absorbers with respect to the hot ICM and cluster substructures.   We also compute the covering fraction of warm gas, discuss the relative distribution of warm gas around gas-rich galaxies, and make comparisons to a cosmological cluster simulation.  The paper is organized as follows.  In \S~\ref{data.sec}, we describe the observational data used in this study, as well as physical properties of the Virgo Cluster.  The overall distribution of warm gas, its covering fraction, and the association of \lya absorbers with Virgo substructures are discussed in \S~\ref{result.sec}.  Finally, \S~\ref{discussion.sec} discusses the results and the origin of warm gas, and \S~\ref{conclusion.sec} briefly summarizes our results. Throughout this paper, the cosmological parameters are assumed to be $\Omega_m=0.3$, $\Omega_{\Lambda}=0.7$, and $H_0= 70 \kms \rm~Mpc^{-1}$.

\section{Data}
\label{data.sec}

We observed 15 QSOs in the background of the Virgo Cluster with COS, which are described in \S~\ref{COS.sec}.  We found 12 QSO spectra could be used to find absorption lines. 
We also searched the literature and found 11 QSO sightlines around the Virgo Cluster that were observed by STIS or GHRS (see \S~\ref{old.sec}). Forty-three \lya absorbers are found along 9 of the COS sightlines and 7 of the STIS or GHRS sightlines and these are described in \S~\ref{absorbers.sec}.
The galaxy catalogs we use in this study and the known physical properties of the Virgo Cluster are noted in \S~\ref{other.sec} and previous studies on the Virgo substructures are summarized in \S~\ref{substructures.sec}.

\subsection{COS Observations}
\label{COS.sec}

We obtained  UV spectra of 15 QSOs behind the Virgo Cluster with COS in $HST$ Cycle 17 (PID 11698; PI Putman). Table~\ref{tab: journal_observations} shows a basic journal of the observations. The QSO sightlines were selected from the catalog of  \citet{Veron-Cetty2006a} to lie within or slightly beyond the virial radius of Virgo and by their FUV brightness from the Galaxy Evolution Explorer  \citep[][]{Martin2005a} source catalog. As we are primarily interested in \lya absorption lines in the Virgo Cluster, all targets were observed only with the G130M grating, using the 1300$\rm~\AA$ central wavelength setting. Our targets were assigned either 1 or 2 orbits of time, based on their FUV brightness. For our 8 single-orbit targets, we employed 2 FP-POS positions to dither the spectra in order to reduce the impact of fixed-pattern and grid-wire flat field effects.  The remaining 7 targets were assigned 2 orbits, and for these we employed all 4 FP-POS positions.

The data were pipeline processed and extracted from the MAST archive using {\it calcos} v2.12. To combine the individual spectra, we employed the coaddition developed for several COS large programs. These routines are fully described in the literature \citep{Meiring2011a,  Thom2011a, Tumlinson2011a}.  Briefly, each exposure in each of the two detector segments is aligned and coadded using a common Milky Way absorption line as a reference (e.g., \SiIII\,1206). Then the two detector segments are co-aligned and combined into a single 1D spectrum. A primary advantage of these coaddition routines is that they operate in count space, and so correctly calculate the poisson errors of the counts in each pixel \citep{Gehrels1986a}. This is beneficial since our data are in the low-count regime ($N < 30$). 

\begin{deluxetable}{llclr}
\tabletypesize{\scriptsize}
\tablecaption{\label{tab: journal_observations}Summary of the COS Observations}
\tablewidth{0pt}
\tablehead{
  \colhead{QSO} &
  \colhead{Short Name} &
  \colhead{$z_{\rm QSO}$} &
  \colhead{$t_{\rm exp}$}\\
  \colhead{} &
  \colhead{} &
  \colhead{} &
  \colhead{(\mbox{s})} 
}
\startdata
SDSSJ120556.08+104253.8 & J1205+1042 & 1.0884  &   4776 \\
SDSSJ120924.07+103612.0$^{\star}$ & J1209+1036 & 0.3949  &   4839 \\
SDSSJ121430.55+082508.1 & J1214+0825 & 0.5854  &   4812 \\
SDSSJ121640.56+071224.3 & J1216+0712 & 0.5865  &   2048 \\
SDSSJ121716.08+080942.0 & J1217+0809 & 0.3428  &   2094 \\
SDSSJ121850.51+101554.2 & J1218+1015 & 0.5424  &   5116 \\
SDSSJ122018.43+064119.6 & J1220+0641 & 0.2864  &   2255 \\
SDSSJ122102.49+155447.0 & J1221+1554 & 0.2294  &   2263 \\
SDSSJ122312.16+095017.7 & J1223+0950 & 0.2771  &   2258 \\
SDSSJ122317.79+092306.9 & J1223+0923 & 0.6815  &   5108 \\
SDSSJ122512.93+121835.6 & J1225+1218 & 0.4118  &   2063 \\
SDSSJ122520.13+084450.7 & J1225+0844 & 0.5350  &   1942 \\
SDSSJ123426.80+072411.3 & J1234+0724 & 0.8439  &   2072 \\
SDSSJ123647.72+060048.4$^{\star}$ & J1236+0600 & 1.2779  &   4743 \\
SDSSJ124035.51+094941.0$^{\star}$ & J1240+0949 & 1.0479  &   5110
\enddata
\tablecomments{$^{\star}$ These sightlines were wiped out by a LLS.}
\end{deluxetable}

\subsection{STIS and GHRS Observations from the Literature}
\label{old.sec}

From the previous STIS and GHRS observations of QSOs in the vicinity of the Virgo Cluster, 11 additional QSO sightlines are selected for this study \citep{Impey1999a,Rosenberg2003a,Penton2004a,Chen2009a,Williger2010a}.   We adopt line identifications and measurements from the references noted. For 3C273,  four \lya lines were consistently detected in the independent studies with GHRS and STIS data \citep{Penton2000a,Williger2010a}.  We, therefore, adopt these four lines. The blended \lya lines at $z=0.005251$ and $z=0.005295$ detected by \citet{Williger2010a} are considered as one line since our COS data cannot resolve these two lines (COS velocity resolution $\sim 15$\kms). The \lya line list of PG1216+069 is adopted from \citet{Chen2009a}.  For the both sightlines, 3C273 and PG1216+069, equivalent widths of the \lya absorbers were not provided in the papers \citep{Chen2009a,Williger2010a}, hence, we re-measure the equivalent widths of these \lya lines from the STIS data.  Details of the spectral processing for the STIS data were described in \citet{Thom2008a}. 

\begin{deluxetable*}{ccrrrrcclr}
\tabletypesize{\scriptsize}
\tablecaption{\label{COSSLs.tab} The COS sightlines with/without \lya absorbers}
\tablewidth{0pt}
\tablehead{
  \colhead{Name } &
  \colhead{RA} &
  \colhead{Decl.} &
  \colhead{$\lambda_{\rm obs} $} &
  \colhead{ $cz$ } &
  \colhead{$W$} &
  \colhead{$\sigma_W$} &
  \colhead{$W_{\rm limit}$$^a$} &
  \colhead{log \nhi$^b$} &
  \colhead{SL$^c$} \\
  \colhead{} &
  \colhead{[J2000,$^{\circ}$]} &
  \colhead{[J2000,$^{\circ}$]} &
  \colhead{[\AA]} &
  \colhead{[$\rm km~s^{-1}$]} &
  \colhead{[$\mA$]} &
  \colhead{[$\mA$]} &
  \colhead{[$\mA$]} &
  \colhead{[$\rm cm^{-2}$]} 
  }
\startdata
J1205+1042 & 181.4837 & 10.7150 & 1219.3033 &  897$^\ast$ & 148 &  22 &  91 & 13.562 &  6.7 \\
            & 181.4837 & 10.7150 & 1219.5564 &  959$^\ast$ & 111 &  21 &  91 & 13.396 &  5.3 \\
            & 181.4837 & 10.7150 & 1223.5604 & 1947 & 140 &  21 &  91 & 13.523 &  6.7 \\
            & 181.4837 & 10.7150 & 1225.1500 & 2339 & 191 &  22 &  91 & 13.719 &  8.7 \\
            & 181.4837 & 10.7150 & 1225.9738 & 2543 & 220 &  27 &  91 & 13.826 &  8.1 \\
 J1214+0825 & 183.6273 &  8.4189 & 1224.0797 & 2075 & 233 &  17 &  60 & 13.865 & 13.7 \\
 J1216+0712 & 184.1690 &  7.2068 & 1223.0385 & 1818 & 370 &  15 &  58 & 14.422 & 24.7 \\
            & 184.1690 &  7.2068 & 1223.4108 & 1910 & 214 &  12 &  58 & 13.807 & 17.8 \\
            & 184.1690 &  7.2068 & 1223.7236 & 1987 &  98 &  20 &  58 & 13.328 &  4.9 \\
            & 184.1690 &  7.2068 & 1224.6844 & 2225 & 409 &  19 &  58 & 14.627 & 21.5 \\
 J1217+0809 & 184.3170 &  8.1617 & 1223.3341 & 1891 & 374 &  13 &  55 & 14.441 & 28.8 \\
            & 184.3170 &  8.1617 & 1224.7227 & 2234 & 444 &  12 &  55 & 14.832 & 37.0 \\
            & 184.3170 &  8.1617 & 1226.1651 & 2590 & 279 &  16 &  55 & 14.031 & 17.4 \\
            & 184.3170 &  8.1617 & 1227.0007 & 2796 & 141 &  13 &  55 & 13.533 & 10.8 \\
 J1218+1015 & 184.7105 & 10.2651 & 1226.9414 & 2782 & 114 &  11 &  53 & 13.406 & 10.4 \\
            & 184.7105 & 10.2651 & 1227.5171 & 2924 &  67 &  12 &  53 & 13.133 &  5.6 \\
 J1220+0641 
            & 185.0768 &  6.6888 & 1222.9626 & 1800 & 138 &  24 &  95 & 13.523 &  5.8 \\
 J1223+0923 & 185.8242 &  9.3853 & 1225.3909 & 2399 &  93 &  11 &  51 & 13.299 &  8.5 \\
 J1225+1218 & 186.3039 & 12.3099 & 1224.8783 & 2272 &  58 &  11 &  60 & 13.064 &  5.3 \\
            & 186.3039 & 12.3099 & 1225.7332 & 2483 & 156 &  18 &  60 & 13.592 &  8.7 \\
 J1225+0844 & 186.3339 &  8.7474 & 1219.1171 &  851$^\ast$ & 523 &  27 &  86 & 15.379 & 19.4 \\
            & 186.3339 &  8.7474 & 1219.7615 & 1010 & 159 &  27 &  86 & 13.602 &  5.9 \\
            & 186.3339 &  8.7474 & 1220.4068 & 1169 & 180 &  34 &  86 & 13.680 &  5.3 \\
            & 186.3339 &  8.7474 & 1221.2135 & 1368 & 270 &  27 &  86 & 14.002 & 10.0 \\
            & 186.3339 &  8.7474 & 1221.6652 & 1479 & 252 &  24 &  86 & 13.934 & 10.5 \\
 J1234+0724 & 188.6117 &  7.4032 & 0.0000 & 0 &  0 &  0 &  76 & 13.201$^\dagger$ &  0.0 \\
 J1221+1554 & 185.2604 & 15.9131 &    0.0000 &    0 &   0 &   0 & 100 & 13.338$^\dagger$ &  0.0 \\
 J1223+0950 & 185.8007 &  9.8383 &    0.0000 &    0 &   0 &   0 &  92 & 13.299$^\dagger$ &  0.0 
\enddata
\tablecomments{ $^a$ $W_{\rm limit}$ is estimated at 1228$\rm~\AA$. \\
$^b$ b is assumed to be 30\kms.\\
$^c$ ${\rm SL}= W/\sigma_W$. \\
$^\dagger W_{\rm limit}$ is converted to column density. \\
$^\ast$ These absorbers in the range $700-1000\kms$ are not used for galaxy correlation and covering fraction analysis (see text).}
\end{deluxetable*}

\subsection{\lya Absorbers}
\label{absorbers.sec}

 In the COS data, we identified a total of 25 \lya absorption lines toward  9 sightlines in the Virgo velocity range ($700 < cz < 3000\kms$, see \S~\ref{other.sec}) above a significance level $W/\sigma_{W} > 4.5$. Three sightlines had no detections over this significance level (see Table~\ref{COSSLs.tab}). We note that the lowest significance level of the absorbers detected is 4.9 which makes our absorber selections robust. The spectra of 3 sightlines, J1209+1036, J1236+0600, and J1240+0949, were wiped out by Lyman limit systems, and thus excluded. 
We do not attempt to detect absorption lines at  velocities $\leq 700\kms$ due to the contamination by the damping wing of the Milky Way \lya absorption.  Through close analysis of the data, we also find completeness is affected for a few sightlines between $700-1000\kms$ due to this damping wing.  We, therefore, present the absorbers within this range in Table~\ref{COSSLs.tab} (marked with $^\ast$), but do not use them in the galaxy correlation and covering fraction analysis.
The 4.5$\sigma$ detection limits in Table~\ref{COSSLs.tab}, $W_{\rm limit}$, are estimated at 1228$\rm~\AA$ for the typical width of a narrow line (0.3$\rm~\AA$).
The column density, \nhi, is estimated from the linear part of the curve of growth assuming $b=30$\kms, as we do not have any saturated lines.   Since we chose the FUV brightest QSOs in the Virgo region, this resulted in a bias toward the southwest half of the cluster as evident in Figure~\ref{ f1.fig}.

In the STIS and GHRS data from the literature, 18 \lya absorption lines toward  7 sightlines and no \lya lines toward  4 sightlines were found in the Virgo velocity range and above the significance level threshold 4.5, as detailed in Table~\ref{oldSLs.tab}.   The data collected from the literature have various sensitivities. The detection limits of \citet{Impey1999a} are provided in the paper.  The detection limits of the sightlines from \citet{Penton2004a} are read from the figures in the paper at 1228$\rm~\AA$ and that of RXJ1230.8+0115 is averaged over the six values from Table 1 of \citet{Rosenberg2003a} as there is no other information available to compute the limit from the paper. We compute the detection limits of 3C273 and PG1216+069 with  the same method used  for the COS data at 1228$\rm~\AA$.

 All 43 \lya absorbers are presented with colored pies in Figure~\ref{ f1.fig}. The size and the number of slices of each pie overlaid on the $\times$ (COS sightline), $+$ (STIS sightline), or $-$ (GHRS sightline) symbols are proportional and equivalent to the number of the \lya absorbers, respectively. The color of each slice corresponds to the velocity of each \lya absorber as presented in the color bar on the right-side.

\begin{deluxetable*}{crrrrrcclrc}
\tabletypesize{\scriptsize}
\tablecaption{\label{oldSLs.tab} The STIS and GHRS sightlines from the literature with/without  \lya absorbers}
\tablewidth{0pt}
\tablehead{
  \colhead{Name } &
  \colhead{RA} &
  \colhead{Dec} &
  \colhead{$\lambda_{\rm obs} $} &
  \colhead{ $cz$ } &
  \colhead{$W$} &
  \colhead{$\sigma_W$} &
  \colhead{$W_{\rm limit}$} &
  \colhead{log\nhi} &
  \colhead{SL} &
  \colhead{Inst.} \\
  \colhead{} &
  \colhead{[J2000,$^{\circ}$]} &
  \colhead{[J2000,$^{\circ}$]} &
  \colhead{[\AA]} &
  \colhead{[$\rm km~s^{-1}$]} &
  \colhead{[$\mA$]} &
  \colhead{[$\mA$]} &
  \colhead{[$\mA$]} &
  \colhead{[$\rm cm^{-2}$]} &
  \colhead{}
  }
\startdata
         3C273$^d$ & 187.2783 &  2.0522 & 1219.7643 & 1013 & 397 &   7 &  13 & 14.559 & 56.7 & STIS\\
                   & 187.2783 &  2.0522 & 1222.1036 & 1582 & 405 &   6 &  13 & 14.607 & 67.5 & --\\
                   & 187.2783 &  2.0522 & 1224.4457 & 2149 &  27 &   4 &  13 & 12.713 &  6.8 & -- \\
                   & 187.2783 &  2.0522 & 1224.9163 & 2276 &  24 &   3 &  13 & 12.654 &  8.0 & -- \\
    PG1216+069$^c$ & 184.8375 &  6.6439 & 1220.0860 & 1090 &  93 &  18 &  62 & 13.299 &  5.2 & STIS \\
                   & 184.8375 &  6.6439 & 1223.3287 & 1890 & 1953 &  31 &  62 & 19.320$^\ddag$ & 63.0 & -- \\
RXJ1230.8+0115$^a$ 
                   & 187.7083 &  1.2560 & 1221.6800 & 1482 & 158 &  14 &  62 & 13.602 & 11.3 & STIS \\
                   & 187.7083 &  1.2560 & 1222.5000 & 1685 & 497 &  13 &  62 & 15.193 & 36.8 & -- \\
                   & 187.7083 &  1.2560 & 1222.6500 & 1721 & 410 &  11 &  62 & 14.637 & 37.3 & -- \\
                   & 187.7083 &  1.2560 & 1223.1100 & 1834 & 115 &  14 &  62 & 13.416 &  8.2 & -- \\
                   & 187.7083 &  1.2560 & 1225.0100 & 2302 & 360 &  17 &  62 & 14.373 & 20.9 & -- \\
    PG1211+143$^b$ & 183.5737 & 14.0533 & 1224.3100 & 2130 & 186 &  19 &  45 & 13.709 &  9.8 & STIS \\
       Ton1542$^b$ & 188.0150 & 20.1581 & 1220.4800 & 1186 & 294 &  56 &  56 & 14.090 &  5.3 & STIS \\
                   & 188.0150 & 20.1581 & 1223.3600 & 1895 & 216 &  42 &  56 & 13.807 &  5.1 & -- \\
                   & 188.0150 & 20.1581 & 1226.0600 & 2563 & 248 &  41 &  56 & 13.924 &  6.0 & -- \\
   PKS1217+023$^e$ & 185.0492 &  2.0617 & 1223.9300 & 2038 & 451 &  88 & 210 & 14.881 &  5.1 & GHRS \\
                   & 185.0492 &  2.0617 & 1224.8300 & 2260 & 648 & 132 & 210 & 16.385 &  4.9 & -- \\
    Q1252+0200$^e$ & 193.8321 &  1.7367 & 1227.2300 & 2853 & 473 &  72 & 270 & 15.027 &  6.6 & GHRS \\
    Q1230+0947$^e$ & 188.3575 &  9.5231 &         0 &    0 &   0 &   0 & 250 & 13.934$^\dagger$ &    0 & GHRS \\
    Q1228+1116$^e$ & 187.7254 & 11.0031 &         0 &    0 &   0 &   0 & 540 & 15.506$^\dagger$ &    0 & GHRS \\
   PKS1252+119$^e$ & 193.6592 & 11.6850 &         0 &    0 &   0 &   0 & 200 & 13.758$^\dagger$ &    0 & GHRS \\
    Q1214+1804$^e$ & 184.2046 & 17.8011 &         0 &    0 &   0 &   0 & 170 & 13.641$^\dagger$ &    0 & GHRS 
\enddata
\tablecomments{ a: \citet{Rosenberg2003a},  b: \citet{Penton2004a}, c: \citet{Chen2009a}, d: \citet{Williger2010a}, e: \citet{Impey1999a} \\
 $^\dagger W_{\rm limit}$ is converted to column density. \\
 $^\ddag$ Adopted from \citet{Tripp2005a}. }
\end{deluxetable*}

\subsection{Other Data and Known Properties of the Virgo Cluster}
\label{other.sec}

In this section, we describe other data  we have collected from the literature in order to aid our analysis.  We use the $ROSAT$ X-ray map from \citet{Bohringer1994a}, shown in Figure~\ref{ f1.fig}. Note that the X-ray data for the southern part, below $\rm Dec\sim5^{\circ}$, is not available and the south-east edge around the virial radius is not reliable as explained in \citet{Bohringer1994a}.  The approximate total mass of the Virgo Cluster, based on X-ray observations, is around $M_{\rm tot} \sim 10^{14-15}~\msun$ \citep{Bohringer1994a,Schindler1999a}.

For the optical galaxy catalog, we use the spectroscopic Sloan Digital Sky Survey (SDSS) DR7 galaxies that have $r_{\rm petro} < 17.7$ \citep[][]{York2000a} in the same velocity range as the \lya absorbers ($700-3000\kms$).  The SDSS galaxies are shown in the right panel of Figure~\ref{ f1.fig}, color-coded by their velocities.  We make this choice because the traditional VCC catalog \citep{Binggeli1985a} only includes a partial list of galaxies in the Virgo substructures.

The \HI galaxy catalog of the \HI Parkes All Sky Survey\citep[HIPASS,][]{Meyer2004a,Wong2006a}, and the Arecibo Legacy Fast ALFA survey\citep[ALFALFA,][]{Giovanelli2007a,Haynes2011a} are also used for comparisons to the absorbers.  HIPASS covers the entire region around the Virgo Cluster, and the ALFALFA galaxy catalog covers the range $\rm 4^{\circ}<decl.<16^{\circ}$.  The $5\sigma$ detection limit is 0.72 Jy\kms~for ALFALFA and 5.6 Jy\kms~for HIPASS \citep{Giovanelli2005a}. Adopting the distance to the Virgo Cluster of 16.5 Mpc \citep{Mei2007a},  the \HI mass limits for ALFALFA and HIPASS are $4.6\times10^7~\msun$ and $3.6\times10^8~\msun$, respectively.

The mean velocity of the Virgo Cluster, $v_{\rm Virgo}$,  is taken to be at 1050\kms~\citep{Binggeli1993a}.  Typically, 3000\kms~is taken as the upper limit on the velocity of Virgo galaxies \citep{Binggeli1993a,Mei2007a}; therefore, we limit our analysis of the galaxies and \lya absorbers in the Virgo Cluster to  $cz < 3000 \kms$.  The center of the Virgo Cluster is assumed to be at M87 (the coordinate from the {\it NASA/IPAC Extragalactic Database}\footnote{http://ned.ipac.caltech.edu}) and the virial radius is adopted to be $1.65h_{2/3}^{-1} \rm~Mpc$ \citep{Mamon2004a}, equivalent to 1.57 Mpc in this study.  However, the Virgo Cluster is an irregular cluster, so a fixed radius is not entirely appropriate (see the discussion in the next section).

\subsection{Virgo Substructures}
\label{substructures.sec}

\begin{deluxetable*}{ccccccc}
\tabletypesize{\scriptsize}
\tablecaption{\label{substructures.tab} Description of the Virgo substructures}
\tablewidth{0pt}
\tablehead{
  \colhead{} &
  \colhead{Cluster A} &
  \colhead{Cluster B} &
  \colhead{W Cloud} &
  \colhead{\Wprime Cloud} &
  \colhead{M Cloud} &
  \colhead{Southern Extension} 
}
\startdata
RA,decl.($^{\circ}$)	& 187.71,12.39	& 187.44,8.00	& 184.88,5.85	& 186.00,6.90	& 183.10,13.20	& 187.50,2.50 \\
$v_{\rm mean} (\kms)$	& 1088$^a$	& 958$^a$	& 2198$^b$	& 1042$^a$	& 2179$^b$	& 1670 \\
$\sigma_{v} (\kms)$	& 593$^a$	& 222$^a$	& 220$^b$	& 253$^a$	& 121$^b$	& 433\\
\# of absorbers$^{\dag}$	& - & 5$^{\ast}$ & 10 & 5$^{\ast}$ & 1 & 7 \\
Kinematics$^{\ddag}$		& Main Cluster	&  -	& possible infall to B &  infall to B & infall to A & Local Supercluster 
\enddata
\tablecomments{ $^{\dag}$ The number of \lya absorbers in spatial and kinematic vicinity of each substructure.\\
$^{\ddag}$ The kinematics of each substructure from the literature (see text). \\
$^{\ast}$ Since there is no clear boundary between Cluster B and the \Wprime cloud, the same 5 absorbers are listed in both places. \\
a: \citet{Mei2007a}, b: \citet{Ftaclas1984a}.}
\end{deluxetable*}

\begin{figure*}
\begin{center}
  \includegraphics[width=2\columnwidth]{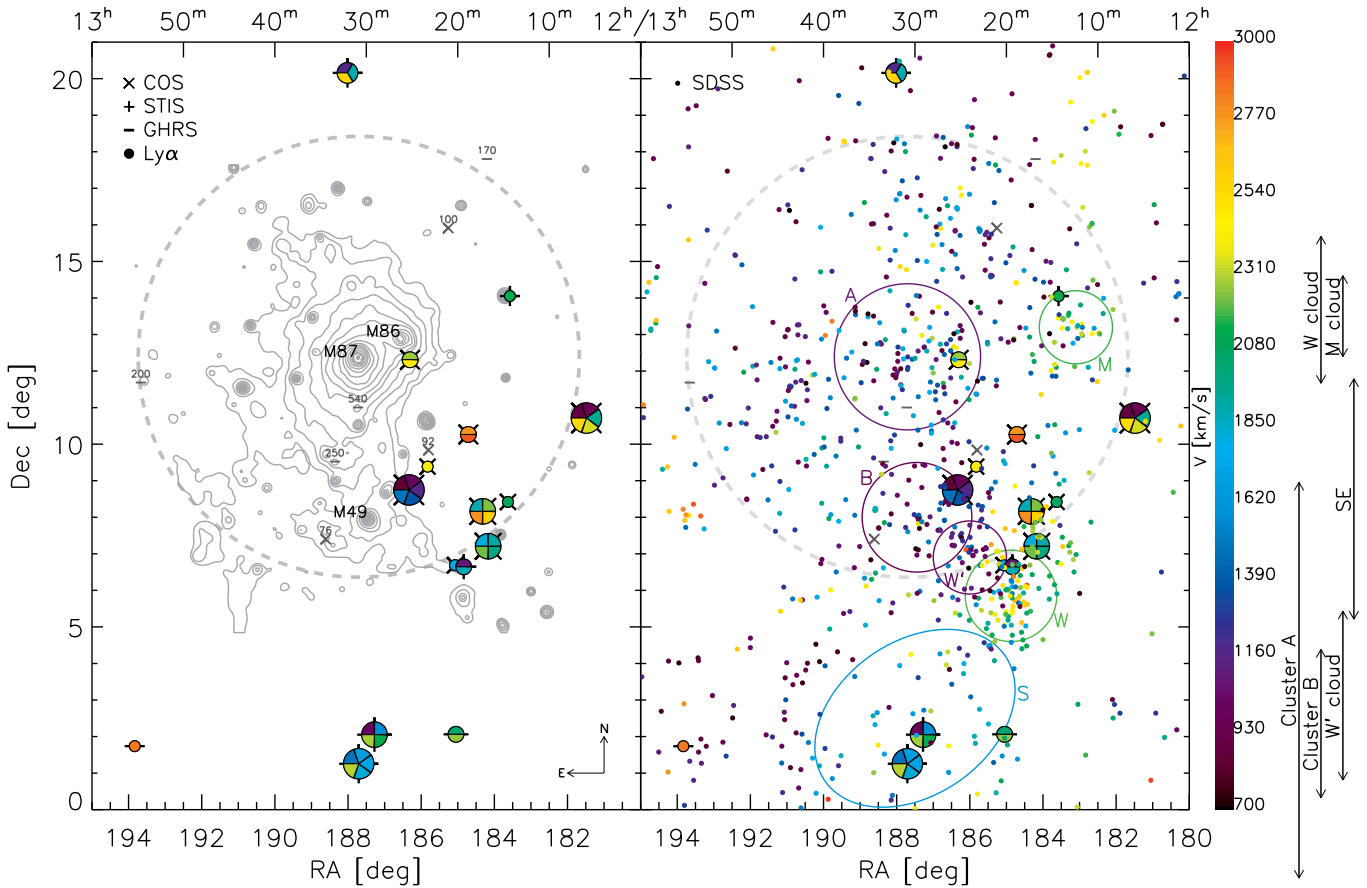}
\caption{Area of the Virgo Cluster and its surroundings with the virial radius of the cluster noted with the large gray dashed circle ($r_{100} = 1.65h^{-1}_{2/3} \rm Mpc$ \citep{Mamon2004a}, equivalent to 1.57 Mpc in this study).  The sightlines used are marked according to the instruments (Cross - COS (this study), Plus - STIS (literature), Bar - GHRS (literature)) and the circle represents the detection of one or more Ly$\alpha$ absorber(s) with the color(s) within the circle representing the velocity(ies) of the absorber(s) (velocity color bar is to the right).   {\it Left}: The QSO sightlines overlaid on the ROSAT X-ray contour map \citep{Bohringer1994a}. The level of the contour corresponds to that of \citet{Bohringer1994a} except that the lowest level here is 5$\sigma$.  Gray crosses and bars with numbers indicate sightlines without \lya detections and the detection limits in \mA. 
{\it Right}: The same as the left panel but with the SDSS galaxies color-coded by their radial velocities. The substructures, A (Cluster A), B (Cluster B), M (M cloud), W (W cloud), and \Wprime(\Wprime cloud) are presented by circles as defined by \citet{Binggeli1987a} and S (Southern Extension), marked with an ellipsoid, is defined by this study. The color of each substructure circle/ellipsoid corresponds to its mean velocity. The velocity ranges of each substructure are also noted by  arrows on the right side of the color bar. }
\label{ f1.fig}
\end{center}
\end{figure*}

 Since we will discuss the sightlines associated with substructures of the Virgo cluster in this paper,  we summarize previous studies on the existence and kinematics of these substructures in this section.  The Virgo cluster is known to have three main substructures which in order of total mass are: Cluster A containing M87, Cluster B which contains M49, and a subcluster of Cluster A containing M86. These are clearly seen as the three clumps in the X-ray map (the left panel of Figure~\ref{ f1.fig}). There are also the M, W, W$^{\prime}$ clouds, and the Southern Extension, in order of coherence of substructures,  behind the Virgo mean velocity \citep{Shapley1929a,Vaucouleurs1961a,Ftaclas1984a,Binggeli1987a}. These substructures are labeled in the right panel of Figure~\ref{ f1.fig}. Note that the circles do not represent exact boundaries but are from estimates by \citet{Binggeli1985a}. The properties of these substructures are also listed in Table~\ref{substructures.tab} and the velocity ranges are noted to the right of the color bar. Previous studies on these substructures include the following.

\begin{itemize}
\item   {\it Clusters A and B.} Cluster A is the main structure as it is the most massive one and Cluster B is the second. Based on kinematics, it was suggested that Cluster B is falling onto Cluster A from the backside \citep{Binggeli1987a,Gavazzi1999a}; however, the latest results concluded that they are at nearly the same distance \citep{Mei2007a}. The over-abundance of the spiral and irregular galaxies with velocities that deviate from a Gaussian distribution in Cluster A indicates that they may originate from an infalled population \citep{Tully1984a,Binggeli1987a,Binggeli1993a}. 

\item {\it Cluster A subcluster.} This subcluster, containing M86, is thought to be $1-2$ Mpc behind the Virgo Cluster \citep{Jerjen2004a,Mei2007a} and M86 has a negative radial velocity of $-244$\kms \citep{Smith2000a}. There are also a few other galaxies around M86 that show negative velocities, at $cz\sim -700\kms$ \citep{Binggeli1993a}. This implies that the subcluster with M86 is infalling to Cluster A from the backside \citep{Binggeli1987a,Binggeli1993a,Bohringer1994a,Jerjen2004a}.  Unfortunately, any \lya absorbers associated with this subcluster, if they exist, cannot be studied in this paper due to the limitation of our data as described in \S~\ref{absorbers.sec}.

\item {\it M and W clouds.} These substructures may be connected to each other \citep{Paturel1979a,Ftaclas1984a}, and falling into the Virgo Cluster together \citep{Yasuda1997a}. However, another study partly disagrees with this suggestion, and argues instead that the Virgo Cluster and the W cloud closely follow the Hubble flow while the M cloud is infalling onto Cluster A \citep{Gavazzi1999a}.

\item {\it \Wprime cloud.}: \citet{Binggeli1993a} proposed that the \Wprime cloud is connected with the W cloud and both are falling into Cluster B, but the distance estimates of the \Wprime cloud make it appear to be more of a localized structure than a connecting bridge \citep{Mei2007a}. The radial velocities of the \Wprime cloud galaxies are consistent with the Virgo inflow model \citep{Mei2007a}. The W and \Wprime clouds are spiral rich \citep{Binggeli1993a}, which supports the idea of both representing an infalling population of galaxies \citep{Tully1984a}.

\item {\it Southern Extension.} There is an extended structure in the south of the Virgo Cluster called the Southern Extension which seems to be a filament in the Local Supercluster \citep{Vaucouleurs1973a,Tully1982a,Hoffman1995a}. This structure is also thought to be infalling to the Virgo Cluster \citep{Binggeli1987a}. The Southern Extension seems to be connected to the W and M clouds (see Figure~\ref{ f5.fig}). This Southern Extension - W cloud - M cloud structure was also noted as a double sheet where galaxies concentrate around 1000\kms~and 2000\kms~in decl.-velocity space \citep[see Figure 6 of ][]{Binggeli1993a}.

We estimate the velocity of the Southern Extension to be 1670$\pm$433\kms~with the {\it biweight} method \citep{Beers1990a} using the galaxies within the `S' ellipsoid in Figure~\ref{ f1.fig}. This was necessary as the definition of the Southern Extension is not clear in previous studies. Note that the velocity dispersion of the Southern Extension galaxies is relatively large because they are not as clustered as the other substructures (although it does show more clustering in the \HI galaxy distribution shown in Figure~\ref{ f5.fig}). 

\end{itemize}

In summary, these substructures are interconnected to each other and generally show kinematics characteristic of infall.  It is important to note that the substructures are actually outside of the virial radius of the Virgo Cluster. The \Wprime cloud is $\sim 6$ Mpc further away from us than the Virgo Cluster \citep{Mei2007a}. The W and M clouds are thought to be twice as distant as the main cluster \citep{Binggeli1987a,Gavazzi1999a}. If we relate  \lya absorbers to these substructures, this implies that many of the absorbers are located beyond the virial radius of the main cluster.  Thus, this paper presents a study of gas both in the Virgo Cluster and in the cosmic filaments feeding into the cluster.

\section{Results}
\label{result.sec}

With our 25 \lya absorbers toward COS sightlines and 18 absorbers toward STIS/GHRS sightlines in the range of $700 < cz < 3000\kms$, we examine the overall distribution of warm gas in the cluster in \S~\ref{dist.sec}, the covering fraction of warm gas in \S~\ref{fcover.sec}, the absorbers associated with the Virgo substructures in \S~\ref{SLs.sec}, the absorbers' relation to the \HI galaxy distribution in \S~\ref{HI.sec}, and the absorbers in the Virgo background in \S~\ref{background.sec}.

\subsection{Large Scale Distribution of \lya Absorbers}
\label{dist.sec}

All QSO sightlines from our COS observations and the literature discussed in \S2 are shown in Figure~\ref{ f1.fig}. We immediately note two visually striking trends: first, the \lya absorbers are more predominant in the outskirts of the cluster and second, the majority of them have radial velocities larger than the mean velocity of the Virgo Cluster (although still lower than 3000\kms, our outer cutoff). The last point needs to be interpreted with some care, as the velocities of the \lya absorbers in this study are constrained to be larger than 700\kms~due to the observational limitation (see \S~\ref{absorbers.sec}).

The \lya absorbers also seem to avoid the X-ray gas region in position-velocity space. Only two absorbers overlap with the contour of the X-ray map.  Even for these two, their velocities are actually more than twice the Virgo mean velocity (2272/2483\kms~vs. 1050\kms). This velocity offset, together with the fact that the X-ray gas usually follows a cluster's potential well, implies that these absorbers are likely to be behind the hot ICM.\footnote{Although it is, of course, not possible to determine a 3D location based on the radial velocity as it cannot be directly translated into distance. For example, although the radial velocity of M86 is negative, the distance to M86 is $1-2$ Mpc larger than the mean Virgo distance.}  Three out of the four sightlines within the X-ray contours do not have any \lya absorbers (one COS sightline and two from GHRS).  This is particularly significant for the COS sightline given its sensitivity ($W_{\rm limit}=76~\mA$).  The two GHRS sightlines, although they have poorer sensitivity ($W_{\rm limit}=250~\mA,~540~\mA$), do at least support the lack of strong absorbers in the X-ray emitting region.  Thus, we conclude that the \lya absorbers avoid the hot ICM in position-velocity space.

\begin{figure}
\begin{center}
  \includegraphics[width=\columnwidth]{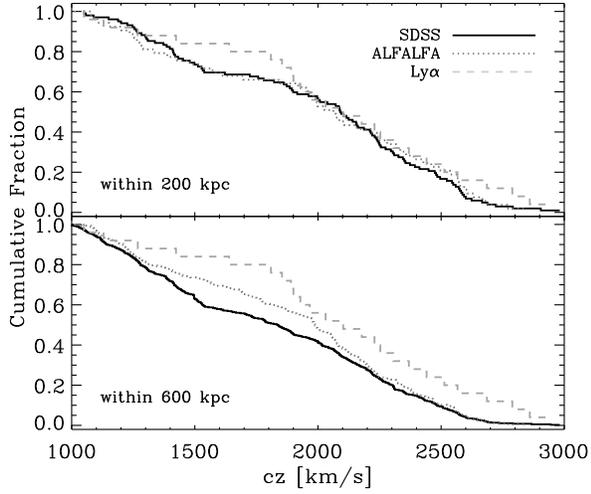}
\caption{The cumulative velocity distribution of the SDSS and ALFALFA galaxies, and the \lya absorbers in the range 1000 - 3000\kms. These data are selected within $4^{\circ}<\rm decl.<16^{\circ}$ due to the limitation of the ALFALFA data.}
\label{ f2.fig}
\end{center}
\end{figure}

Next, we address the relation between the absorbers and the galaxy catalogs.  To do this, we compute the velocity distributions of the SDSS and ALFALFA galaxies within a radius of 200, 300, 400, 500, and 600 kpc around each sightline with the \lya absorbers within $\rm 4^{\circ} < decl. < 16^{\circ}$. To test if the velocity distributions of each of these populations are consistent with being drawn from the same parent distribution, in Table~\ref{ks.tab}, we present the result of the Kolmogorov-Smirnov (K-S) test in the velocity range $1000 - 3000\kms$. The cumulative distributions of the velocities of the galaxies in the 200 and 600 kpc cases are illustrated in Figure~\ref{ f2.fig}. 

Based on the computed K-S probability, we cannot differentiate the ALFALFA (\ion{H}{1}) galaxies from the \lya absorbers with a high significance level.  Hence, we conclude that cold gas (\ion{H}{1}) coexists with warm gas (\lya absorbers) on all scales. This is consistent with the finding of a positive association between \lya absorbers and gas-rich galaxies by \cite{Ryan-Weber2006a,Pierleoni2008a}.
 However, for the case of the SDSS (optical) galaxies and \lya absorbers (warm gas), as the search radius of galaxies around each sightline increases, we can reject the hypothesis that they are drawn from the same parent population with a significance level $>94\%$ except the case for 200 and 300 kpc. This is mostly because as the search radius grows, the optical galaxies in the cluster center are included which do not typically have warm or cold components due to environmental effects.  As the search radius decreases, we approach a more localized scale (e.g., a galaxy halo) and the distributions of SDSS, ALFALFA, and \lya absorbers become more indistinguishable.

\begin{figure}
\begin{center}
  \includegraphics[width=\columnwidth]{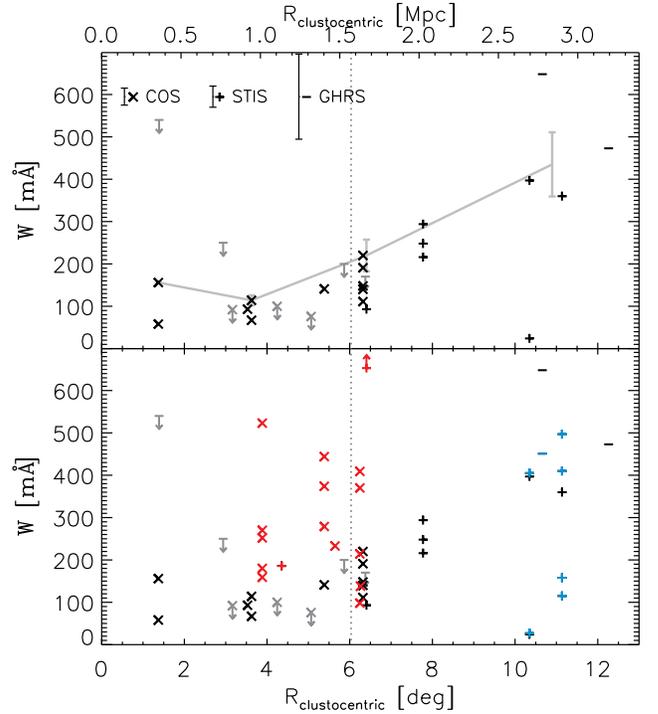}
\caption{Cluster impact parameter (clustocentric radius in projection from M87) and equivalent widths ($W$) of the \lya absorbers for the sightlines from COS ($\times$), STIS (+), and GHRS (--). The top panel only includes the sightlines not related to the substructures and the bottom shows all the sightlines. The gray solid line in the top panel illustrates the weighted least square regression of the maximum W of each sightline alongside its mean error in four radius bins with a bin size 3.1$^{\circ}$. Red indicates the sightlines in the region of the W, \Wprime, M cloud, and Cluster B, and blue represents sightlines in the Southern Extension. Gray symbols with downward arrows present the detection limits of each sightline without any \lya absorbers. The vertical dotted line presents the virial radius as presented in Figure~\ref{ f1.fig}. The mean error bars of the absorbers from each instrument are illustrated in the top-left with the legend. The red plus with an upward arrow actually has a $W$ of $1953~\mA$ (sub-DLA system, see text).  }
\label{ f3.fig}
\end{center}
\end{figure}

We investigate the distribution of  \lya absorbers as a function of a projected radius from the cluster center in Figure~\ref{ f3.fig}. We divide the \lya absorbers into two classes, depending on whether their position and velocity can be associated with the Virgo substructures (red and blue symbols) or not (black symbols).  
We consider an absorber to be associated with a substructure  if it is spatially consistent with being within it and within 500\kms~of the substructure's velocity (see Table~\ref{substructures.tab} and Figure~\ref{ f1.fig}).  As described further in \S~\ref{SLs.sec}, the following sightlines in Figure~\ref{ f1.fig} have substructure related absorbers:  J1214+0825, J1216+0712, J1217+0809, J1220+0641, and PG1216+069 with the W/\Wprime cloud, J1223+0923 and J1225+0844 with the Cluster B/\Wprime Cloud, PG1211+143 in the M cloud, and 3C273, RXJ1230.8+0115, and PKS1217+023 in the Southern Extension (`S' ellipsoid).  
All other absorbers (black symbols) are within the velocity range of the Virgo and do not have nearby clearly related substructures. We note that all the strong absorbers ($> 160~\mA$) within the virial radius in projection coincide with the Virgo substructures in position-velocity space (bottom panel of Figure~\ref{ f3.fig}).  When we exclude the \lya absorbers associated with the substructures, there are only weak absorption lines within the virial radius in projection (top panel of Figure~\ref{ f3.fig}). Regardless of whether these weak absorbers actually reside in the virial radius or are due to a projection effect, strong absorbers are not found in the virial radius and upper limits on line strengths for the non-detection sightlines also support this argument.  It is also notable that in the top panel of Figure~\ref{ f3.fig} without the substructure absorbers, there appears to be a gradual trend of increasing line strengths with a projected radius. The gray-thick line in the top panel illustrates a weighted linear regression of the maximum line strength of each sightline in 4 bins.  Overall, we find a suppression of the \lya absorbers within the virial radius in projection and the presence of the strong absorbers with the substructures and outside the virial radius. 

\begin{deluxetable*}{ccccc}
\tabletypesize{\scriptsize}
\tablecaption{\label{ks.tab} The K-S probability for the SDSS and ALFALFA galaxies versus the \lya absorbers around each sightline within various radii in the range $4^\circ<\rm decl.<16^\circ$.}
\tablewidth{0pt}
\tablehead{
  \colhead{ } &
  \colhead{ $P_{\rm K-S}$ } &
    \colhead{$N_{\rm SDSS}$ } &
  \colhead{ $P_{\rm K-S}$ } &
    \colhead{$N_{\rm ALFALFA}$ }   \vspace{1mm} \\
  \colhead{ } &
  \colhead{SDSS-\lya} &
   \colhead{\#} &
  \colhead{ALFALFA-\lya} &
   \colhead{\#} 
  }
\startdata
$1000 < cz < 3000 \kms$, $R<$200 kpc & 0.699 & 102 & 0.750 & 53 \\
$1000 < cz < 3000 \kms$, $R<$300 kpc & 0.124 & 194 & 0.525 & 90 \\
$1000 < cz < 3000 \kms$, $R<$400 kpc & 0.056 & 278 & 0.477 & 124 \\
$1000 < cz < 3000 \kms$, $R<$500 kpc & 0.030 & 329 & 0.245 & 147 \\
$1000 < cz < 3000 \kms$, $R<$600 kpc & 0.030 & 373 &0.307 & 167 
\enddata
\tablecomments{See the text for the details of the ALFALFA and SDSS galaxy selections.}
\end{deluxetable*}

\begin{figure}
\begin{center}
  \includegraphics[width=\columnwidth]{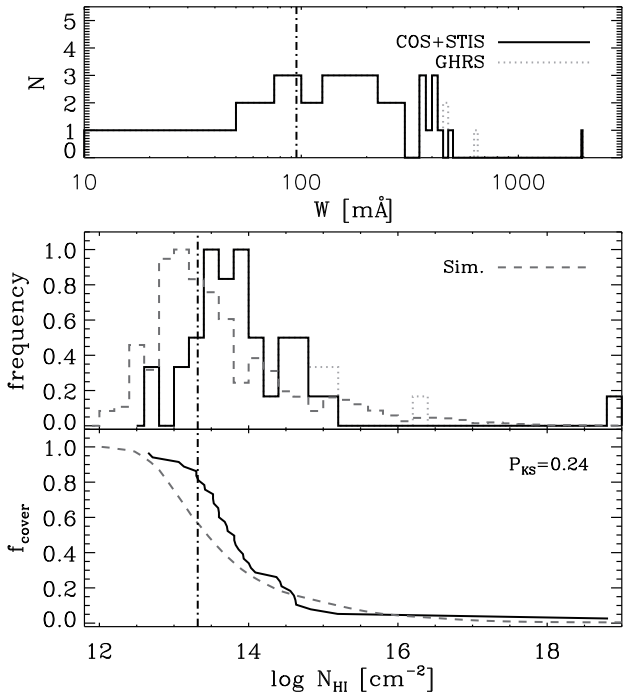}
\caption{The distribution of equivalent widths $W$, column densities, and covering fraction.  The top panel shows the distribution of observed equivalent widths ($W$) where black shows the absorbers from COS and STIS sightlines and the light-gray dotted line shows the GHRS data.  The middle and bottom panels show the column densities and covering fraction for the observations (same line coding) and the cosmological simulation (dashed gray line). Vertical dot-dashed lines indicate the completeness cut at $95~\mA$ ($\nhi=10^{13.3}\cm-2$) for the COS and STIS data completeness in the K-S test. 
 In the bottom panel, the covering fraction of the observed data is normalized to 0.81 which is estimated by counting sightlines with/without \lya detections stronger than $95~\mA$. The left-side of the vertical dot-dashed line for the observed covering fraction is incomplete due to sensitivity issues (see the text). On top-right of the bottom panel, the K-S probability between the observation and simulation is noted.}
\label{ f4.fig}
\end{center}
\end{figure}

\subsection{The Covering Fraction of Warm Gas}
\label{fcover.sec}

\begin{deluxetable*}{ccccc}
\tabletypesize{\scriptsize}
\tablecaption{\label{fcover.tab} The covering fraction of warm gas in the velocity range $1000-3000\kms$ in different radius bins}
\tablewidth{0pt}
\tablehead{
  \colhead{$W$ } &
  \colhead{log \nhi } &
  \colhead{} &
  \colhead{$f_{\rm cover}$} &
  \colhead{}\\
  \colhead{[$\mA$]} &
  \colhead{[$\rm cm^{-2}$]} &
  \colhead{$0-1R_{\rm vir}$} &
  \colhead{$0-2R_{\rm vir}$} &
  \colhead{$1-2R_{\rm vir}$}
  }
\startdata \vspace{1mm}
$ > 65$ &13.123  & 1.00$_{-0.15}$(6/6,1.00) & 1.00$_{-0.09}$(11/11,1.00)  & 1.00$_{-0.17}$(5/5,1.00) \\ \vspace{1mm}
$>100$ & 13.338 & 0.60$_{-0.16}^{+0.13}$(6/10,0.62) & 0.76$_{-0.12}^{+0.08}$(13/17,0.70) & 1.00$_{-0.14}$(7/7,0.73) \\ \vspace{1mm}
$>150$ & 13.572 & 0.50$_{-0.15}^{+0.15}$(5/10,0.55) & 0.65$_{-0.12}^{+0.10}$(11/17,0.68) & 0.86$_{-0.18}^{+0.08}$(6/7,0.72) \\ \vspace{1mm}
$>200$ & 13.758 & 0.27$_{-0.10}^{+0.15}$(3/11,0.18) & 0.47$_{-0.11}^{+0.11}$(9/19,0.44) & 0.75$_{-0.17}^{+0.11}$(6/8,0.52) \\ \vspace{1mm}
$>300$ & 14.109 & 0.08$_{-0.05}^{+0.12}$(1/12,0.02) & 0.29$_{-0.09}^{+0.11}$(6/21,0.17) & 0.56$_{-0.16}^{+0.15}$(5/9,0.22) \\  
 \hline \vspace{1mm} 
$>65$ & ($3700-5700 \kms$) & 0.33$_{-0.14}^{+0.19}$(2/6,0.08) && \\ \vspace{1mm}
$>65$ & ($10000-12000 \kms$) & 0.33$_{-0.14}^{+0.19}$(2/6,0.45) &&\\ \vspace{1mm}
$>65$ & ($15000-17000 \kms$) & 0.33$_{-0.14}^{+0.19}$(2/6,0.48) &&
\enddata
\tablecomments{In parentheses, following the covering fraction and its 1$\sigma$ lower and upper limits from likelihood functions,  $(N_{\rm sightlines}$ with detections) / (Total $N_{\rm sightlines}$) and the covering fraction from the Voronoi tessellation method (see text) are listed. \\
The bottom three rows show the covering fraction of gas ($>65~\mA$) in the background of the Virgo Cluster with the velocity range noted. }
\end{deluxetable*}

We examine the covering fraction of warm gas by counting the number of the sightlines with and without \lya absorbers within $2R_{\rm vir}$ and in the velocity range $1000-3000\kms$.  The estimated total volume of this region is about $\pi (2R_{\rm vir})^2\times33 \rm~Mpc^3$, using the Virgo distance of 16.5 Mpc for the {\it x-} and {\it y-}axes and the distances to known Virgo sub-structures and Hubble flow considerations for the {\it z-}axis (see \S\ref{substructures.sec}).   A covering fraction can be estimated with the ratio of the number of sightlines with a detected absorber to the total number of sightlines.  When doing this calculation, we use five different  thresholds  ($W > 65, ~100, ~150, ~200, ~300~\mA$), and ignore the  sightlines for which the detection limit is larger than the specified threshold (since we cannot determine if that sightline has a detection at the specified threshold or not).  Note that the estimated covering fraction is limited by the non-uniform sightline distribution. 

In Table~\ref{fcover.tab}, we present the covering fractions in different radius bins $0-1R_{\rm vir}$, $0-2R_{\rm vir}$, and $1-2R_{\rm vir}$. The gas covering fraction, $\nhi > 10^{13.123}\cm-2 (=65~\mA)$, is unity within $2R_{\rm vir}$ of the Virgo Cluster. 
The covering fraction of gas with higher column density is less than unity, but becomes higher outside of the virial radius ($1-2R_{\rm vir}$). The covering fraction is computed for the background (i.e., for a part of the spectra with significantly higher velocities, as detailed in Table~\ref{fcover.tab}) in order to compare with non-cluster environments. We find that the covering fraction of warm gas in the background is less than half that in the environment of the Virgo Cluster. 

As previously noted, the sightlines in the south-west of the cluster are more clustered than those in other regions. In order to alleviate any possible spatial bias, we also use Voronoi tessellation and compute the area of a polygon around each sightline as an independent way to estimate the covering fraction. The ratio of the sum of the area of the polygons with detections to the total area in $1R_{\rm vir}$ and $2R_{\rm vir}$ is defined to be the covering fraction. Hence, the Voronoi tessellation method assumes that the correlation length of a \lya absorbing cloud is larger than the minimum transverse distance between the sightlines.  The estimates from this Voronoi tessellation method are also listed in Table~\ref{fcover.tab}, in brackets following the results from the counting method. We see that the results from both counting sightlines and Voronoi tessellation do not differ from each other significantly and show the same trend.

The distributions of the \lya absorbers' equivalent widths and column densities are shown in Figure~\ref{ f4.fig} and compared to a cosmological simulation of a cluster carried out with the adaptive mesh refinement code {\it ENZO} \citep{Bryan1999a,Norman1999a,O'Shea2004a}. The cluster simulation has the same physics but has a higher resolution than the run of \citet{Tonnesen2007a}. The simulation has a particle mass of $\sim2 \times 10^8$~\msun, a spatial resolution of 2 kpc, and includes a prescription for radiative cooling, star formation and feedback, as well as an ionizing background from \citet{Haardt2001a}. The simulated cluster has $M_{200}=8\times10^{14}~\msun$ and a gas temperature of $\sim 5$ keV.  \HI fractions are computed by post-processing with {\it CLOUDY} \citep{Ferland1998a}. To do the comparison, we projected the \HI column density along one dimension of a 1000 Mpc$^3$ box (10~Mpc each side) which is comparable to (but somewhat smaller than) the volume from our observations.  The \HI gas in the simulated cluster peaks at $10^{13} \cm-2$ and the covering fraction is nearly unity when $\nhi \sim 10^{12.5}\cm-2$. There are varying detection limits for each observational sightline, but we draw a line at 95~\mA~($ \nhi = 10^{13.3}\cm-2$) as a completeness limit, which is the maximum detection limit of the COS and STIS sightlines with \lya detections. In the top and middle panel of Figure~\ref{ f4.fig}, at the cut, shown by a vertical dot-dashed line, there is a clear drop in the number of absorbers.  In the bottom panel, we also show the result of applying the K-S test, finding that $P_{\rm K-S}=0.24$  at $\nhi>10^{13.3}$\cm-2 ($W=95~\mA$, the completeness limit of our data). This indicates that the gas covering fraction from the observation and simulation does not significantly differ from each other. Note that the simulated covering fraction is integrated through the box while the observed absorbers are broken up into multiples. A more careful comparison will be performed in a future paper.

\subsection{Absorbers Related to the Virgo Substructures}
\label{SLs.sec}

 In this section, we discuss the likely associations of the absorbers noted with the color symbols in Figure~\ref{ f3.fig} to the infalling substructures described in \S~\ref{substructures.sec}.  The \lya absorbers are considered associated with the substructures if close in position-velocity space as noted in the last paragraph of \S~\ref{dist.sec} and shown in Figure~\ref{ f1.fig}.  As mentioned previously, all the strong absorbers ($>$160~\mA) in the virial radius are found toward the sightlines penetrating through the substructures.

There are many \lya absorbers concentrated in the southwest region where the W, \Wprime cloud, and Cluster B are connected to each other. We also see an overdensity of SDSS galaxies with  similar velocity ranges to the substructures in that region (see the right panel of Figure~\ref{ f1.fig}).  The sightline J1225+0844 is near Cluster B and the \Wprime cloud and has 5 \lya absorbers in the velocity range of the galaxies in these substructures. The sightlines J1214+0825, J1216+0712, J1217+0809, J1220+0641, and PG1216+069,  are near the W and \Wprime Clouds, and have many \lya absorbers that mostly coincide with the W cloud in position-velocity space.

 There are also several \lya absorbers which seem to concur with the M cloud and the Southern Extension.
The sightline PG1211+143 in the M cloud, in Figure~\ref{ f1.fig}, has only one \lya absorption line and this is also near the presence of an \HI filament in its velocity range (as discussed in \S~\ref{HI.sec}).
The three sightlines in the Southern Extension, 3C273, RXJ1230.8+0115, and PKS1217+023 (in the `S' ellipsoid of the right panel of Figure~\ref{ f1.fig}), also show \lya absorbers coinciding with the filamentary structure of galaxies in position-velocity space.  This was also noted by previous studies \citep{Bahcall1991a,Morris1991a,Morris1993a,Salpeter1995a}.

The remaining absorbers are not considered  part of known substructures.
The sightline J1225+1218, near the subcluster containing M86, has two \lya absorbers at $cz = 2272$ and $2483\kms$. However, this subcluster is infalling to Cluster A from backside and the galaxies in it exhibit negative velocities as discussed in \S~\ref{substructures.sec}.  Thus, these absorbers are not likely to be associated with this substructure, but more likely in the background of the hot ICM as discussed in \S~\ref{dist.sec}.

The remaining sightlines with \lya absorbers show various environments. Although there are only a handful of galaxies around the sightline J1205+1042 (rightmost symbol in Figure~\ref{ f1.fig}) and Ton1542 (top symbol in Figure~\ref{ f1.fig}), they have 5 and 3 absorbers respectively. The sightline J1234+0723 (gray X labeled 76) is in Cluster B and has no absorbers, while J1218+1015 has two low column density absorbers ($10^{13.1}$ and $10^{13.4}\cm-2$) at high velocities (2782 \& 2924\kms, colored with orange and red) with no galaxies around it in position and velocity space.

\begin{figure*}
\begin{center}
  \includegraphics[width=2\columnwidth]{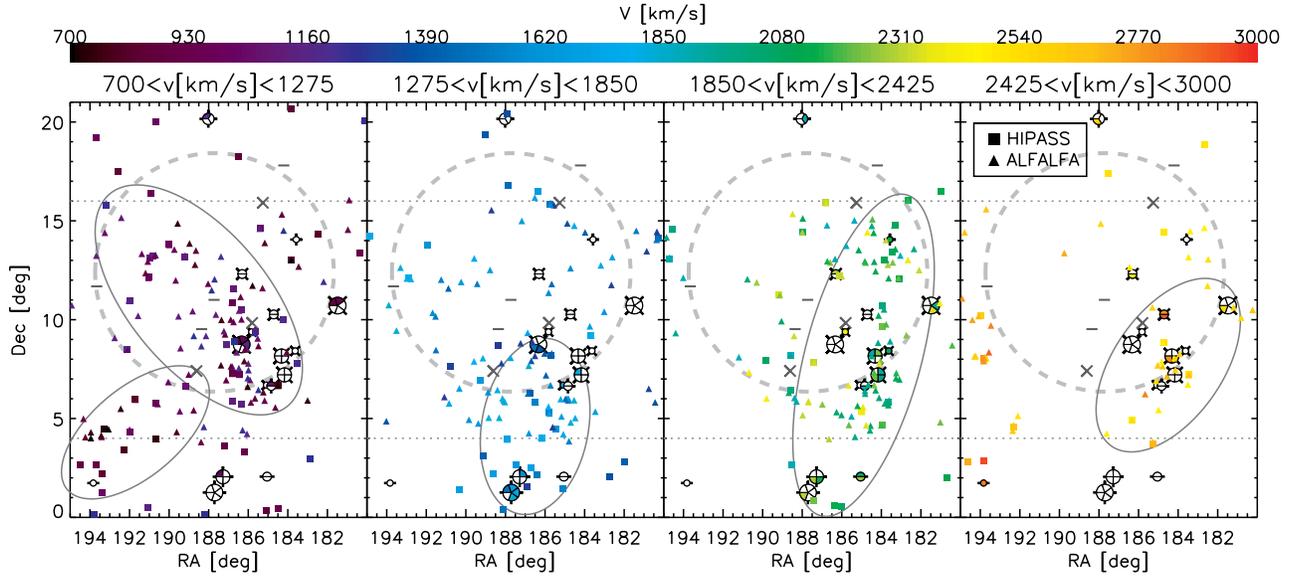}
\caption{The same area and velocity coverage as Figure~\ref{ f1.fig}, but with the ALFALFA and HIPASS galaxies and separated into four velocity bins as noted on the top of each panel. Possible filamentary structures are indicated by gray ellipsoids (see text).}
\label{ f5.fig}
\end{center}
\end{figure*}

\subsection{Ly$\alpha$ Absorbers and \HI Galaxies}
\label{HI.sec}

\HI observations of the galaxies in the Virgo Cluster are presented in several pointed studies \citep[i.e.,][]{Hoffman1989a,Gavazzi2005a,Chung2009a}, as well as in \HI surveys which cover a large fraction of the sky \citep{Wong2006a, Haynes2011a}. The HIPASS survey covers the entire region around the Virgo Cluster and the ALFALFA survey observed most of our region of interest.  We plot the HIPASS and ALFALFA galaxies in different velocity bins with the \lya absorbers to investigate the cold and warm gas distributions in Figure~\ref{ f5.fig}. The ALFALFA survey region is illustrated between the horizontal dotted lines and is also evident with the larger abundance of \HI galaxies due to the survey depth differences (see \S~\ref{other.sec}).

In the left-most panel of Figure~\ref{ f5.fig}, around the sightline J1225+0844 (with five \lya absorbers near M49 in Figure~\ref{ f1.fig}), the overdensity of \HI detections is clearly visible and highlighted by the large gray ellipsoid. This is where Cluster B and the \Wprime cloud meet. The \HI overdensity near J1225+0844 seems to be parallel to the edge of the X-ray contour of Figure~\ref{ f1.fig}. As discussed in \citet{Tully1984a}, this overabundance of gas-rich galaxies could originate from an infalling population. The smaller gray ellipse to the south-east of this panel contains an overabundance of HI galaxies that is also seen in the SDSS galaxy distribution (see the right panel of Figure~\ref{ f1.fig}).  Though not discussed in their study, this structure can also be partially seen in the HI maps of \citet{Popping2011a}. There are no \lya absorbers detected here, but the  GHRS sightline from the literature has poor sensitivity ($W_{\rm limit}=270~\mA$). Future work may detect \lya absorbers in this region.

In the second and third panels, we see abundant \HI detections that link the Southern Extension to the W and M clouds.
 The majority of the \lya absorbers in this filamentary structure lie at a similar velocity to the \HI filament. The connecting structure of the Southern Extension, W, and M cloud was noted in previous studies (as mentioned in \S~\ref{substructures.sec}), but is seen here for the first time in both cold and warm gas. In the last panel, although there are not many detections, the majority of the \HI sources and the \lya absorbers are found around the W cloud.

\subsection{Gaseous Structures behind The Virgo Cluster}
\label{background.sec}

\begin{deluxetable*}{crrrcccccrc}
\tabletypesize{\scriptsize}
\tablecaption{\label{bkg.tab} The \lya absorbers in the Virgo background}
\tablewidth{0pt}
\tablehead{
  \colhead{Name } &
  \colhead{RA} &
  \colhead{Decl.} &
  \colhead{$\lambda_{\rm obs} $} &
  \colhead{ $cz$ } &
  \colhead{$W$} &
  \colhead{$\sigma_W$} &
  \colhead{$W_{\rm limit}$} &
  \colhead{log\nhi} &
  \colhead{SL} &
  \colhead{Inst.} \\
  \colhead{} &
  \colhead{[J2000,$^{\circ}$]} &
  \colhead{[J2000,$^{\circ}$]} &
  \colhead{[\AA]} &
  \colhead{[$\rm km~s^{-1}$]} &
  \colhead{[$\mA$]} &
  \colhead{[$\mA$]} &
  \colhead{[$\mA$]} &
  \colhead{[$\rm cm^{-2}$]} &
  \colhead{}
  }
\startdata
        J1205+1042 & 181.4837 & 10.7150 & 1235.3704 &  4862 & 388 &  20 &  91 & 14.510 & 19.4 & COS \\
        J1214+0825 & 183.6273 &  8.4189 & 1231.7817 &  3976 & 297 &  20 &  60 & 14.100 & 14.9 & COS \\
        J1216+0712 & 184.1690 &  7.2068 & 1230.8713 &  3776 &  61 &   9 &  58 & 13.094 &  6.8 & COS \\
                   & 184.1690 &  7.2068 & 1231.0527 &  3796 &  70 &   9 &  58 & 13.162 &  7.8 & -- \\
                   & 184.1690 &  7.2068 & 1231.2832 &  3853 & 131 &  14 &  58 & 13.484 &  9.4 & -- \\
        J1217+0809 & 184.3170 &  8.1617 & 1231.9110 &  4008 & 211 &  14 &  55 & 13.797 & 15.1 & COS \\
                   & 184.3170 &  8.1617 & 1238.8798 &  5720 & 148 &  16 &  55 & 13.562 &  9.3 & -- \\
                   & 184.3170 &  8.1617 & 1239.6354 &  5924 & 145 &  11 &  55 & 13.543 & 13.2 & -- \\
        J1218+1015 & 184.7105 & 10.2651 & 1239.6635 &  5921 & 300 &  13 &  53 & 14.109 & 23.1 & COS \\
        J1220+0641 & 185.0768 &  6.6888 & 1232.3699 &  4121 & 163 &  22 &  95 & 13.621 &  7.4 & COS \\
    PG1216+069$^c$ & 184.8375 &  6.6439 & 1231.0341 &  3792 & 296 &  17 &  62 & 14.100 & 17.4 & STIS \\
                   & 184.8375 &  6.6439 & 1233.8179 &  4478 &  44 &   9 &  62 & 12.938 &  4.9 & -- 
\enddata
\tablecomments{ c: references from Table~\ref{oldSLs.tab}}
\end{deluxetable*}

\begin{figure*}
\begin{center}
  \includegraphics[width=2\columnwidth]{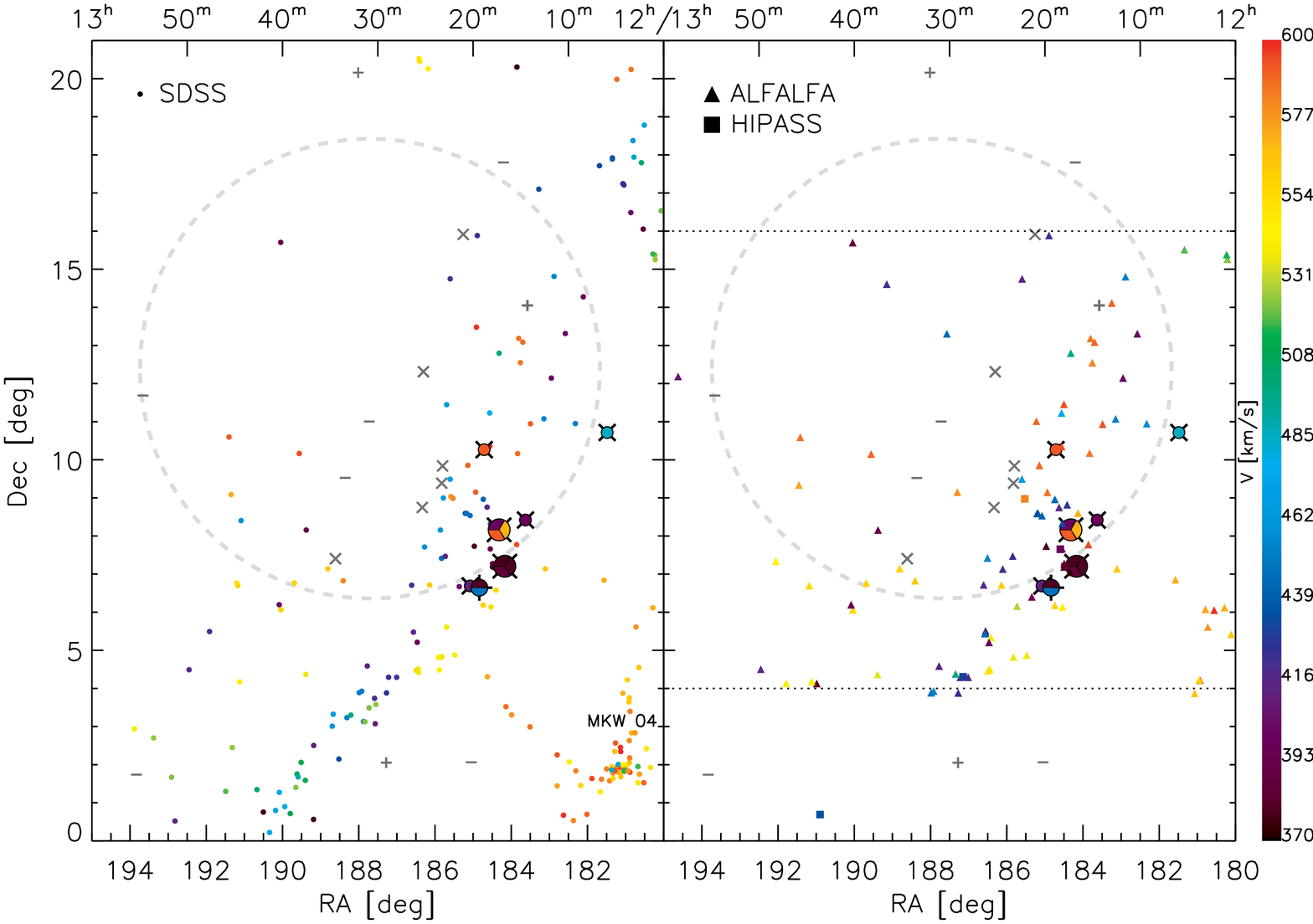}
\caption{The \lya absorbers in the background of the Virgo Cluster with the SDSS (left) and ALFALFA/HIPASS (right) galaxies. Symbols for the sightlines are the same as Figure~\ref{ f1.fig} and \ref{ f5.fig}. The ALFALFA galaxies are only available between the two dotted lines in the right panel. All galaxies and absorbers are color coded by their velocities as noted in the bar on the right side. }
\label{ f6.fig}
\end{center}
\end{figure*}

Our multiple sightlines in the Virgo Cluster allow us to probe structures beyond the Virgo environment as well. The \lya absorbers in the velocity range of $3700-6000\kms$ are listed in Table~\ref{bkg.tab} and they are presented with the SDSS, ALFALFA, and HIPASS galaxies in Figure~\ref{ f6.fig}. There are 10 \lya absorbers toward the COS sightlines and 2 absorbers toward the STIS sightlines. A filamentary structure from the south-east to the north-west is evident in Figure~\ref{ f6.fig} and the \lya absorbers coexist with this filament in position-velocity space.
Another filament of warm gas and galaxies seems to be evident in a branch from the galaxy cluster MKW 04 to the three \lya absorbers at $\sim$5800\kms (orange and red). There are no \lya absorbers in the range $3000-3700\kms$. This absorber gap is in agreement with the existence of the gap in the galaxy distribution \citep{Binggeli1993a}.

The covering fraction of the \lya absorbers  in this background region is $0.33_{-0.14}^{+0.19}$, as determined by counting the sightlines and $0.08$ by Voronoi tessellation ($W > 65~\mA$, $3700<cz<5700\kms$ to have the same velocity range). This is lower than in (and around) the Virgo Cluster and the same as the other two comparison samples in the velocity ranges $10000-12000\kms$ and $15000-17000\kms$ (Table~\ref{fcover.tab}). These ranges are arbitrarily chosen to compare absorber statistics over a similar velocity interval in non-cluster environments.

\section{Discussion}
\label{discussion.sec}

Warm gas in the environment of a galaxy cluster can be interpreted in the context of predictions from cosmological simulations of cluster formation and its relation to the other observed phases of a cluster.  In this section, we discuss the distribution and origin of the \lya absorbers. We also discuss the comparison of our results to simulation results and  the relation of the absorbers to hot X-ray gas, galaxies (all and only the gas-rich), and  Virgo substructures.

\subsection{Diffuse Warm Gas and the Virgo Cluster}
\label{overall.sec}

We detect  warm gas surrounding the Virgo Cluster, but few absorbers are apparently within the virial radius of the cluster.  This is inferred because most of the absorbers are at velocities higher than the mean velocity of the Virgo Cluster and/or can be associated with the infalling substructures. They are also largely detected beyond the X-ray contours, and those within the X-ray contours in projection have higher velocities than the Virgo mean velocity.  In contrast to within the cluster, the outskirts of the cluster show an abundance of the absorbers relative to the background indicating warm circumcluster gas is abundant. This can be interpreted in the context of warm gas coming into a cluster and being shock heated, as predicted for gas coming into a massive dark matter halo such as Virgo \citep{White1978a}.    It is also consistent with the hot ICM stripping any cold/warm gas from galaxies and that material rapidly being integrated in the hot ICM \citep{Chung2007a,Tonnesen2007a}. Indeed, the majority of warm gas in simulations is not found in virialized structures, consistent with our finding of abundant warm circumcluster gas beyond the virial radius \citep{Dave2001a,Dave2010a,Shull2011a}. 

The high covering fraction found for  warm gas also points toward abundant circumcluster gas (see Table 6).
The covering fraction of warm gas with $\nhi > 10^{13.1} \cm-2$ ($W>65\mA$) within $2R_{\rm vir}$ in projection is unity.  As mentioned above, this is thought to largely represent surrounding gas, and not gas within the virial radius, as our velocity cut at 3000\kms~includes not only Virgo, but the substructures that are twice as distant as the Virgo Cluster (see \S~\ref{substructures.sec}). This projection effect is partially evident when examining the trend toward lowering the covering fraction from $0-1R_{\rm vir}$ to $1-2R_{\rm vir}$ for the higher column density absorbers.  This clearly shows that the higher column density absorbers are preferentially found in the outskirts of the galaxy cluster, and may be partially due to blending of multiple absorbers along the line of sight.   Using background velocities as discussed in \S~\ref{background.sec}, the covering fraction is lower, showing again the prevalence of warm circumcluster gas relative to non-cluster environments.  Figure~\ref{ f3.fig} also shows that the high column density absorbers within the virial radius in projection  coexist with the substructures.  

We find  agreement between a simulation of a galaxy cluster and our absorber statistics within our completeness limit (Figure 4).  This suggests that the continued feeding of galaxy clusters found in simulations is consistent with the observations. A more thorough investigation will be completed in a future paper. Although the K-S probability shows consistency, we see indications that there are a greater number of absorbers with  densities between $10^{13.3-14.5}\cm-2$ in the observations. This may be due to either the volume in the observations being deeper or the Virgo Cluster's irregular morphology with the multiple infalling substructures not being closely matched by the simulation.  On the other hand, there are more high column absorbers $>10^{15.5}\cm-2$ in the simulation.  Since the column densities in the simulation are obtained by integrating through the box, while the observations are broken up into multiple absorbers, some of the discrepancy may be resolved in future comparisons of individual absorbers in the simulation.   
While the gas in a galaxy sized halo has been studied extensively in simulations \citep{Dekel2009a,Keres2009a,Faucher-Giguere2011a,Kimm2011a}, detailed studies of circumcluster gas are just beginning.

\subsection{What gives Rise to the \lya Absorbers?}
\label{what.sec}

As shown, many \lya absorbers are concurrent with the Virgo substructures.  In particular, the strong absorbers coincide with the substructures. The K-S test shows that the distributions of the \HI galaxies and the \lya absorbers are not distinguishable.  The coincidence of the Virgo infalling substructures traced in cold gas from the \HI observations and warm gas by the \lya absorbers may represent the accretion of both galaxies and intergalactic gas into the cluster and the growth of its diffuse gaseous, stellar, and dark matter components. This is in  agreement with simulations that find the vast majority of \lya absorbers are associated with the overdensities comparable to typical filaments \citep{Dave2001a,Dave2010a,Shull2011a}.  

\begin{deluxetable*}{ccccc}
\tabletypesize{\scriptsize}
\tablecaption{\label{OVI.tab} \OVI absorbers in the FUSE data published in  \citet{Danforth2008a} }
\tablewidth{0pt}
\tablehead{
  \colhead{Name } &
  \colhead{$cz$} &
  \colhead{$W(1032)\pm \sigma_{W}$} &
  \colhead{$W(1038)\pm \sigma_{W}$} &
  \colhead{$N_{\rm OVI}$}\\
  \colhead{} &
  \colhead{[$\rm km~s^{-1}$]} &
  \colhead{[\mA]} &
  \colhead{[\mA]} &
  \colhead{[\cm-2]} 
  }
\startdata
         3C273  & 1017  & 23$\pm$3 & . &13.33$\pm$0.11 \\
PG1211+143  & 2133  &  . & 32$\pm$5 & 13.78$\pm$0.15
\enddata
\end{deluxetable*}

  We see evidence for a connected filament between the Southern Extension - W cloud - M cloud in the \HI galaxy distribution (Figure~\ref{ f5.fig}) and the multiple \lya absorbers in this study. This may  be partially represented by the $\sim2000\kms$ component of the double sheet structure detected in galaxies \citep[][discussed in \S~\ref{substructures.sec}]{Binggeli1993a}.  Although the absorbers toward J1218+1015 and J1205+1042 have relatively few galaxies around them, they are in position-velocity proximity to the Southern Extension - W cloud - M cloud filament.  The 7 \lya absorbers along these two sightlines are relatively weak ($<220~\mA$), which is consistent with them tracing a continuation of this filament, rather than gas in a yet unseen galaxy halo \citep{Prochaska2011a}. Considering the transverse distance between PG1211+143 and RXJ1230.8+0115 (the two extremes of the filament), the gaseous filament extends for at least 3.6 Mpc.  It most likely extends even further given the previous work has inferred the existence of a coherent 20 Mpc (in depth) gaseous structure for the sightlines 3C273 and RXJ1230.8+0115 in the Southern Extension \citep{Penton2002a}.

In these infalling filaments, shocks will heat gas \citep{Dave2001a,Birnboim2003a}, and this  $T=10^{5-6}\rm~K$ gas can potentially be traced by \OVI absorption lines \citep{Shull2003a,Furlanetto2005a}.  We found four sightlines 3C273, RXJ1230.8+0115, PG1211+143, and PG1216+069 observed with $FUSE$ \citep[][]{Sembach2001a,Danforth2006a,Danforth2008a,Tripp2008a}, and two of them, 3C273 (in the Southern Extension) and PG1211+143 (in the M cloud), show  \OVI absorption lines with a significance level $> 4.5$ at the velocities of the \lya absorbers at $cz=1013\kms$ and $cz= 2130\kms$ (see Table 8). The origin of the \OVI absorbers is still debated, so cannot be definitively linked to collisional ionization due to infalling cluster gas. Additional observations and comparisons to simulations are needed.

\subsection{IGM vs. Galactic Origins}

The concurrence of the \HI emission and \lya absorbers can be attributed to IGM gas in a filament that will most likely feed galaxies, or to cold or warm gas being stripped/ejected from galaxies. In either case, we expect to see both warm gas and cold gas in proximity, although most likely with less gas in galaxies in the latter case.  This coincidence of the \HI emission and \lya absorbers also echoes the notion that optical and radio surveys are only seeing the tip of the iceberg. There is more mass, more diffuse warm gas, surrounding those dense regions. 

Previous studies on \lya absorbers did not consider the influence of an ``active" environment, which is the case for the Virgo Cluster. Galaxies begin to lose their gas just outside the virial radius in the Virgo Cluster \citep{Chung2007a}. The gas tails caused by environmental stripping can extend up to $100-500$ kpc \citep{Oosterloo2005a,Haynes2007a,Koopmann2008a} and can produce a \lya absorption line \citep{Tonnesen2010a}. However, the stripped gas will be heated by thermal conduction from the ICM, as well as by turbulence and shocks \citep{Tonnesen2010a}, and may not last long.  Our finding that the \lya absorbers avoid the observed X-ray gas may indicate this heating is relatively rapid in the region where the hot ICM is present. \lya absorbers can only arise from stripped gas around the virial radius where gas stripping begins, but no hot ICM exists. \lya absorbers well beyond the virial radius and the hot ICM (or in the substructures) may not be influenced by environmental gas stripping, but could still partially originate from galaxy interactions and feedback processes.  

Since the substructures are gas-rich in galaxies and absorbers, many of the absorbers may represent gas in a filament that will eventually feed galaxies.
Due to the lack of distance and 3D kinematic information for the \lya absorbers, one promising method to discriminate infalling gas from  gas that has originated from a galaxy is to look at metallicity.  At high and intermediate redshifts most of the accreting IGM is likely to have low metallicities \citep{Faucher-Giguere2011a,Fumagalli2011a,Kimm2011a}, and, thus far, at $z=0$ the IGM metallicity in local filaments still appears to be low \citep[$Z\lesssim0.1Z_{\odot}$]{Ferrara2000a, Danforth2005a, Barai2011a}. To help determine the metallicity of warm gas, there are a few metal lines, such as \SiIII [1206], \SiII [1260], \OI[1302], and  \CII[1334] in the spectral range of our COS data ($1150-1450\rm~\AA$);  however, the column densities of our \lya absorbers are too low to detect any corresponding metal lines. In previous observations, the metallicity of two \lya absorbers at 1013\kms and 1582\kms~with $\nhi>10^{14}\cm-2$ toward 3C273 turned out to be low ($\rm [O/H]\ge-2.0$ and $\rm [C/H] = -1.2$; \cite{Tripp2002a}).  In addition, one of the two \lya absorbers toward PG1216+069 at 1890\kms, in the proximity of the W cloud, was found to be a pristine sub-DLA system without any bright galaxies close by \citep{Tripp2005a}. This implies the presence of the relatively un-enriched IGM and is consistent with much of the gas in the substructure filaments being galactic fuel rather than galactic waste.

\section{Conclusion}
\label{conclusion.sec}

The distribution of warm gas in a galaxy cluster is studied for the first time in this paper with observations of multiple QSO sightlines.  We performed COS observations of bright background QSOs and found 25 \lya absorbers ($\nhi = 10^{13.1-15.4}\cm-2$) toward 9 of 12 QSO sightlines in and around the virial radius of the Virgo Cluster in the velocity range $700 < cz < 3000\kms$. Eighteen \lya absorbers ($\nhi = 10^{12.7-19.3}\cm-2$)  from STIS and GHRS observations are added to our sample by searching the literature and using a more extended region around the Virgo Cluster. The absorbers' overall distribution, covering fraction, and relation to substructures/filaments and gas-rich (\HI) galaxies are investigated and compared to a cosmological simulation. Our findings can be summarized as the following:

\begin{enumerate}
\item Warm gas prefers the outskirts of the cluster and avoids the region of the hot ICM.  Also, there is an indication for an increase in the strength of the \lya absorbers with  cluster impact parameter when only the non-substructure absorbers are considered. While the \lya absorbers in the cluster center are suppressed, the high column density absorbers ($W>160~\mA$, $\nhi > 10^{13.6}\cm-2$) are coincident with the Virgo substructures or reside in the outskirts.
\item Warm gas preferentially found in the filamentary substructures is coincident with an abundance of \HI galaxies.  The distributions of the \HI galaxies and the \lya absorbers show no statistical difference. This coexistence is consistent with the interpretation that we are seeing the flow of gas in different phases into the Virgo cluster along the filaments.
\item The covering fraction of warm gas increases with a projected radius and is consistent with a high resolution cosmological grid simulation of a cluster.  The warm gas covering fraction in the non-cluster environments ($f_{\rm cover}=0.33$ at $\nhi > 10^{13.1}\cm-2$) is lower than that in the cluster ($f_{\rm cover}=1$ at $\nhi > 10^{13.1}\cm-2$) implying the presence of abundant warm circumcluster gas.  
\end{enumerate}

Our findings of warm gas being dominant in the outskirts of the cluster and overall avoiding the hot ICM are consistent with the expected properties of gas flowing into a massive dark matter halo and being  shock heated.  It is also consistent with the stripped gas from galaxies being rapidly heated as it enters the virial radius of a cluster.   Since we find gas in the regions with abundant \HI galaxies and where there is not active stripping going on, much of the warm gas may represent potential galactic fuel, rather than waste.   Metallicity can be an important future diagnostic of whether the material has been ejected from the nearby galaxies.  There is some evidence from the literature that one of the Virgo substructure filaments represents low metallicity warm gas.  More detailed comparisons to simulations and studies of individual galaxies near sightlines will also be important to clarify the origin of the gas.

\acknowledgments
We thank Joel Bregman, Stephanie Tonnesen, Ryan M. Joung, Jacqueline van Gorkom, Jessica Rosenberg, and Michael Anderson for useful discussions.   We acknowledge funding from NASA grant HST-GO-11698, the Luce Foundation, and NSF CAREER grant AST-0904059.

\clearpage
\bibliographystyle{apj}
\bibliography{ms}

\begin{thebibliography}{113}
\expandafter\ifx\csname natexlab\endcsname\relax\def\natexlab#1{#1}\fi

\bibitem[{Bahcall {et~al.}(1991)Bahcall, Jannuzi, Schneider, Hartig, Bohlin, \&
  Junkkarinen}]{Bahcall1991a}
Bahcall, J., Jannuzi, B., Schneider, D., Hartig, G., Bohlin, R., \&
  Junkkarinen, V. 1991, ApJ, 377, L5

\bibitem[{Bahcall \& Spitzer(1969)}]{Bahcall1969a}
Bahcall, J.~N., \& Spitzer, L. 1969, \apj, 156, L63

\bibitem[{Barai {et~al.}(2011)Barai, Martel, \& Germain}]{Barai2011a}
Barai, P., Martel, H., \& Germain, J. 2011, ApJ, 727, 54

\bibitem[{Beers {et~al.}(1990)Beers, Flynn, \& Gebhardt}]{Beers1990a}
Beers, T.~C., Flynn, K., \& Gebhardt, K. 1990, \aj, 100, 32

\bibitem[{Binggeli {et~al.}(1993)Binggeli, Popescu, \& Tammann}]{Binggeli1993a}
Binggeli, B., Popescu, C.~C., \& Tammann, G.~A. 1993, \aaps, 98, 275

\bibitem[{Binggeli {et~al.}(1985)Binggeli, Sandage, \& Tammann}]{Binggeli1985a}
Binggeli, B., Sandage, A., \& Tammann, G.~A. 1985, \aj, 90, 1681

\bibitem[{Binggeli {et~al.}(1987)Binggeli, Tammann, \& Sandage}]{Binggeli1987a}
Binggeli, B., Tammann, G.~A., \& Sandage, A. 1987, \aj, 94, 251

\bibitem[{Birnboim \& Dekel(2003)}]{Birnboim2003a}
Birnboim, Y., \& Dekel, A. 2003, \mnras, 345, 349

\bibitem[{B{\"o}hringer {et~al.}(1994)B{\"o}hringer, Briel, Schwarz, Voges,
  Hartner, \& Tr{\"u}mper}]{Bohringer1994a}
B{\"o}hringer, H., Briel, U.~G., Schwarz, R.~A., Voges, W., Hartner, G., \&
  Tr{\"u}mper, J. 1994, \nat, 368, 828

\bibitem[{{Bryan}(1999)}]{Bryan1999a}
{Bryan}, G.~L. 1999, Comput.~Sci.~Eng., Vol.~1, No.~2, p.~46 - 53, 1, 46

\bibitem[{Burns {et~al.}(2010)Burns, Skillman, \& O'Shea}]{Burns2010a}
Burns, J.~O., Skillman, S.~W., \& O'Shea, B.~W. 2010, \apj, 721, 1105

\bibitem[{Carilli \& van Gorkom(1992)}]{Carilli1992a}
Carilli, C.~L., \& van Gorkom, J.~H. 1992, \apj, 399, 373

\bibitem[{Cen \& Ostriker(1999)}]{Cen1999a}
Cen, R., \& Ostriker, J.~P. 1999, ApJ, 514, 1

\bibitem[{Chen {et~al.}(1998)Chen, Lanzetta, Webb, \& Barcons}]{Chen1998a}
Chen, H.-W., Lanzetta, K.~M., Webb, J.~K., \& Barcons, X. 1998, ApJ, 498, 77

\bibitem[{Chen {et~al.}(2001)Chen, Lanzetta, Webb, \& Barcons}]{Chen2001a}
---. 2001, \apj, 559, 654

\bibitem[{Chen \& Mulchaey(2009)}]{Chen2009a}
Chen, H.-W., \& Mulchaey, J.~S. 2009, ApJ, 701, 1219

\bibitem[{Chung {et~al.}(2007)Chung, van~Gorkom, Kenney, \&
  Vollmer}]{Chung2007a}
Chung, A., van~Gorkom, J., Kenney, J., \& Vollmer, B. 2007, ApJ, 659, L115

\bibitem[{Chung {et~al.}(2009)Chung, van Gorkom, Kenney, Crowl, \&
  Vollmer}]{Chung2009a}
Chung, A., van Gorkom, J.~H., Kenney, J.~D.~P., Crowl, H., \& Vollmer, B. 2009,
  \aj, 138, 1741

\bibitem[{C{\^o}t{\'e} {et~al.}(2004)C{\^o}t{\'e}, Blakeslee, Ferrarese,
  Jord{\'a}n, Mei, Merritt, Milosavljevi{\'c}, Peng, Tonry, \&
  West}]{Cote2004a}
C{\^o}t{\'e}, P., {et~al.} 2004, \apjs, 153, 223

\bibitem[{C{\^o}t{\'e} {et~al.}(2005)C{\^o}t{\'e}, Wyse, Carignan, Freeman, \&
  Broadhurst}]{Cote2005a}
C{\^o}t{\'e}, S., Wyse, R. F.~G., Carignan, C., Freeman, K.~C., \& Broadhurst,
  T. 2005, ApJ, 618, 178

\bibitem[{Danforth \& Shull(2005)}]{Danforth2005a}
Danforth, C., \& Shull, J. 2005, ApJ, 624, 555

\bibitem[{Danforth \& Shull(2008)}]{Danforth2008a}
Danforth, C.~W., \& Shull, J.~M. 2008, ApJ, 679, 194

\bibitem[{Danforth {et~al.}(2006)Danforth, Shull, Rosenberg, \&
  Stocke}]{Danforth2006a}
Danforth, C.~W., Shull, J.~M., Rosenberg, J.~L., \& Stocke, J.~T. 2006, ApJ,
  640, 716

\bibitem[{Dav{\'e} {et~al.}(1999)Dav{\'e}, Hernquist, Katz, \&
  Weinberg}]{Dave1999a}
Dav{\'e}, R., Hernquist, L., Katz, N., \& Weinberg, D.~H. 1999, ApJ, 511, 521

\bibitem[{Dav{\'e} {et~al.}(2010)Dav{\'e}, Oppenheimer, Katz, Kollmeier, \&
  Weinberg}]{Dave2010a}
Dav{\'e}, R., Oppenheimer, B.~D., Katz, N., Kollmeier, J.~A., \& Weinberg,
  D.~H. 2010, \mnras, 408, 2051

\bibitem[{Dav{\'e} {et~al.}(2001)Dav{\'e}, Cen, Ostriker, Bryan, Hernquist,
  Katz, Weinberg, Norman, \& O'Shea}]{Dave2001a}
Dav{\'e}, R., {et~al.} 2001, \apj, 552, 473

\bibitem[{Davies {et~al.}(2010)Davies, Baes, Bendo, Bianchi, Bomans, Boselli,
  Clemens, Corbelli, Cortese, Dariush, de~Looze, di~Serego~Alighieri, Fadda,
  Fritz, Garcia-Appadoo, Gavazzi, Giovanardi, Grossi, Hughes, Hunt, Jones,
  Madden, Pierini, Pohlen, Sabatini, Smith, Verstappen, Vlahakis, Xilouris, \&
  Zibetti}]{Davies2010a}
Davies, J.~I., {et~al.} 2010, A{\&}A, 518, L48

\bibitem[{de~Vaucouleurs(1961)}]{Vaucouleurs1961a}
de~Vaucouleurs, G. 1961, \apjs, 6, 213

\bibitem[{de~Vaucouleurs \& de~Vaucouluers(1973)}]{Vaucouleurs1973a}
de~Vaucouleurs, G., \& de~Vaucouluers, A. 1973, A{\&}A, 28, 109

\bibitem[{Dekel {et~al.}(2009)Dekel, Birnboim, Engel, Freundlich, Goerdt,
  Mumcuoglu, Neistein, Pichon, Teyssier, \& Zinger}]{Dekel2009a}
Dekel, A., {et~al.} 2009, \nat, 457, 451

\bibitem[{Dinshaw {et~al.}(1997)Dinshaw, Weymann, Impey, Foltz, Morris, \&
  Ake}]{Dinshaw1997a}
Dinshaw, N., Weymann, R.~J., Impey, C.~D., Foltz, C.~B., Morris, S.~L., \& Ake,
  T. 1997, \apj, 491, 45

\bibitem[{Ettori {et~al.}(2004)Ettori, Borgani, Moscardini, Murante, Tozzi,
  Diaferio, Dolag, Springel, Tormen, \& Tornatore}]{Ettori2004a}
Ettori, S., {et~al.} 2004, \mnras, 354, 111

\bibitem[{Faucher-Gigu{\`e}re \& Kere{\v s}(2011)}]{Faucher-Giguere2011a}
Faucher-Gigu{\`e}re, C.-A., \& Kere{\v s}, D. 2011, \mnras, 412, L118

\bibitem[{{Ferland} {et~al.}(1998){Ferland}, {Korista}, {Verner}, {Ferguson},
  {Kingdon}, \& {Verner}}]{Ferland1998a}
{Ferland}, G.~J., {Korista}, K.~T., {Verner}, D.~A., {Ferguson}, J.~W.,
  {Kingdon}, J.~B., \& {Verner}, E.~M. 1998, \pasp, 110, 761

\bibitem[{Ferrara {et~al.}(2000)Ferrara, Pettini, \& Shchekinov}]{Ferrara2000a}
Ferrara, A., Pettini, M., \& Shchekinov, Y. 2000, \mnras, 319, 539

\bibitem[{Ftaclas {et~al.}(1984)Ftaclas, Struble, \& Fanelli}]{Ftaclas1984a}
Ftaclas, C., Struble, M.~F., \& Fanelli, M.~N. 1984, \apj, 282, 19

\bibitem[{Fumagalli {et~al.}(2011)Fumagalli, Prochaska, Kasen, Dekel, Ceverino,
  \& Primack}]{Fumagalli2011a}
Fumagalli, M., Prochaska, J.~X., Kasen, D., Dekel, A., Ceverino, D., \&
  Primack, J.~R. 2011, \mnras, 1589

\bibitem[{Furlanetto {et~al.}(2005)Furlanetto, Phillips, \&
  Kamionkowski}]{Furlanetto2005a}
Furlanetto, S.~R., Phillips, L.~A., \& Kamionkowski, M. 2005, \mnras, 359, 295

\bibitem[{Gavazzi {et~al.}(1999)Gavazzi, Boselli, Scodeggio, Pierini, \&
  Belsole}]{Gavazzi1999a}
Gavazzi, G., Boselli, A., Scodeggio, M., Pierini, D., \& Belsole, E. 1999,
  \mnras, 304, 595

\bibitem[{Gavazzi {et~al.}(2005)Gavazzi, Boselli, van Driel, \&
  O'neil}]{Gavazzi2005a}
Gavazzi, G., Boselli, A., van Driel, W., \& O'neil, K. 2005, A{\&}A, 429, 439

\bibitem[{Gehrels(1986)}]{Gehrels1986a}
Gehrels, N. 1986, \apj, 303, 336

\bibitem[{Giovanelli {et~al.}(2005)Giovanelli, Haynes, Kent, Perillat,
  Saintonge, Brosch, Catinella, Hoffman, Stierwalt, Spekkens, Lerner, Masters,
  Momjian, Rosenberg, Springob, Boselli, Charmandaris, Darling, Davies,
  Garcia~Lambas, Gavazzi, Giovanardi, Hardy, Hunt, Iovino, Karachentsev,
  Karachentseva, Koopmann, Marinoni, Minchin, Muller, Putman, Pantoja, Salzer,
  Scodeggio, Skillman, Solanes, Valotto, van Driel, \& van
  Zee}]{Giovanelli2005a}
Giovanelli, R., {et~al.} 2005, \aj, 130, 2598

\bibitem[{Giovanelli {et~al.}(2007)Giovanelli, Haynes, Kent, Saintonge,
  Stierwalt, Altaf, Balonek, Brosch, Brown, Catinella, Furniss, Goldstein,
  Hoffman, Koopmann, Kornreich, Mahmood, Martin, Masters, Mitschang, Momjian,
  Nair, Rosenberg, \& Walsh}]{Giovanelli2007a}
---. 2007, \aj, 133, 2569

\bibitem[{Green {et~al.}(2012)Green, Froning, Osterman, Ebbets, Heap,
  Leitherer, Linsky, Savage, Sembach, Michael~Shull, Siegmund, Snow, Spencer,
  Alan~Stern, Stocke, Welsh, B{\'e}land, Burgh, Danforth, France, Keeney,
  McPhate, Penton, Andrews, Brownsberger, Morse, \& Wilkinson}]{Green2012a}
Green, J.~C., {et~al.} 2012, ApJ, 744, 60

\bibitem[{{Haardt} \& {Madau}(2001)}]{Haardt2001a}
{Haardt}, F., \& {Madau}, P. 2001, in Clusters of Galaxies and the High
  Redshift Universe Observed in X-rays, ed. {D.~M.~Neumann \& J.~T.~V.~Tran}

\bibitem[{Haynes {et~al.}(2007)Haynes, Giovanelli, \& Kent}]{Haynes2007a}
Haynes, M.~P., Giovanelli, R., \& Kent, B.~R. 2007, ApJ, 665, L19

\bibitem[{Haynes {et~al.}(2011)Haynes, Giovanelli, Martin, Hess, Saintonge,
  Adams, Hallenbeck, Hoffman, Huang, Kent, Koopmann, Papastergis, Stierwalt,
  Balonek, Craig, Higdon, Kornreich, Miller, O'Donoghue, Olowin, Rosenberg,
  Spekkens, Troischt, \& Wilcots}]{Haynes2011a}
Haynes, M.~P., {et~al.} 2011, \aj, 142, 170

\bibitem[{Hoffman {et~al.}(1995)Hoffman, Lewis, \& Salpeter}]{Hoffman1995a}
Hoffman, G.~L., Lewis, B.~M., \& Salpeter, E.~E. 1995, \apj, 441, 28

\bibitem[{Hoffman {et~al.}(1989)Hoffman, Williams, Lewis, Helou, \&
  Salpeter}]{Hoffman1989a}
Hoffman, G.~L., Williams, B.~M., Lewis, B.~M., Helou, G., \& Salpeter, E.~E.
  1989, \apjs, 69, 65

\bibitem[{Impey {et~al.}(1999)Impey, Petry, \& Flint}]{Impey1999a}
Impey, C.~D., Petry, C.~E., \& Flint, K.~P. 1999, \apj, 524, 536

\bibitem[{Jerjen {et~al.}(2004)Jerjen, Binggeli, \& Barazza}]{Jerjen2004a}
Jerjen, H., Binggeli, B., \& Barazza, F.~D. 2004, \aj, 127, 771

\bibitem[{Kere{\v s} {et~al.}(2009)Kere{\v s}, Katz, Fardal, Dav{\'e}, \&
  Weinberg}]{Keres2009a}
Kere{\v s}, D., Katz, N., Fardal, M., Dav{\'e}, R., \& Weinberg, D.~H. 2009,
  \mnras, 395, 160

\bibitem[{Kimm {et~al.}(2011)Kimm, Slyz, Devriendt, \& Pichon}]{Kimm2011a}
Kimm, T., Slyz, A., Devriendt, J., \& Pichon, C. 2011, \mnras, 413, L51

\bibitem[{Koekemoer {et~al.}(1998)Koekemoer, O'Dea, Baum, Sarazin, Owen, \&
  Ledlow}]{Koekemoer1998a}
Koekemoer, A.~M., O'Dea, C.~P., Baum, S.~A., Sarazin, C.~L., Owen, F.~N., \&
  Ledlow, M.~J. 1998, ApJ, 508, 608

\bibitem[{Koopmann {et~al.}(2008)Koopmann, Giovanelli, Haynes, Kent, Balonek,
  Brosch, Higdon, Salzer, \& Spector}]{Koopmann2008a}
Koopmann, R.~A., {et~al.} 2008, ApJ, 682, L85

\bibitem[{Kravtsov {et~al.}(2005)Kravtsov, Nagai, \& Vikhlinin}]{Kravtsov2005a}
Kravtsov, A.~V., Nagai, D., \& Vikhlinin, A.~A. 2005, \apj, 625, 588

\bibitem[{Lanzetta {et~al.}(1995)Lanzetta, Bowen, Tytler, \&
  Webb}]{Lanzetta1995a}
Lanzetta, K.~M., Bowen, D.~V., Tytler, D., \& Webb, J.~K. 1995, \apj, 442, 538

\bibitem[{Lanzetta {et~al.}(1996)Lanzetta, Webb, \& Barcons}]{Lanzetta1996a}
Lanzetta, K.~M., Webb, J.~K., \& Barcons, X. 1996, \apjl, 456, L17

\bibitem[{Lehner {et~al.}(2007)Lehner, Savage, Richter, Sembach, Tripp, \&
  Wakker}]{Lehner2007a}
Lehner, N., Savage, B.~D., Richter, P., Sembach, K.~R., Tripp, T.~M., \&
  Wakker, B.~P. 2007, ApJ, 658, 680

\bibitem[{Loken {et~al.}(2002)Loken, Norman, Nelson, Burns, Bryan, \&
  Motl}]{Loken2002a}
Loken, C., Norman, M.~L., Nelson, E., Burns, J., Bryan, G.~L., \& Motl, P.
  2002, \apj, 579, 571

\bibitem[{Lopez {et~al.}(2008)Lopez, Barrientos, Lira, Padilla, Gilbank,
  Gladders, Maza, Tejos, Vidal, \& Yee}]{Lopez2008a}
Lopez, S., {et~al.} 2008, ApJ, 679, 1144

\bibitem[{Mamon {et~al.}(2004)Mamon, Sanchis, Salvador-Sol{\'e}, \&
  Solanes}]{Mamon2004a}
Mamon, G.~A., Sanchis, T., Salvador-Sol{\'e}, E., \& Solanes, J.~M. 2004, \aap,
  414, 445

\bibitem[{Martin {et~al.}(2005)Martin, Fanson, Schiminovich, Morrissey,
  Friedman, Barlow, Conrow, Grange, Jelinsky, Milliard, Siegmund, Bianchi,
  Byun, Donas, Forster, Heckman, Lee, Madore, Malina, Neff, Rich, Small,
  Surber, Szalay, Welsh, \& Wyder}]{Martin2005a}
Martin, D.~C., {et~al.} 2005, ApJ, 619, L1

\bibitem[{Mei {et~al.}(2007)Mei, Blakeslee, C{\^o}t{\'e}, Tonry, West,
  Ferrarese, Jord{\'a}n, Peng, Anthony, \& Merritt}]{Mei2007a}
Mei, S., {et~al.} 2007, ApJ, 655, 144

\bibitem[{Meiring {et~al.}(2011)Meiring, Tripp, Prochaska, Tumlinson, Werk,
  Jenkins, Thom, O'Meara, \& Sembach}]{Meiring2011a}
Meiring, J.~D., {et~al.} 2011, ApJ, 732, 35

\bibitem[{Meyer {et~al.}(2004)Meyer, Zwaan, Webster, Staveley-Smith,
  Ryan-Weber, Drinkwater, Barnes, Howlett, Kilborn, Stevens, Waugh, Pierce,
  Bhathal, De~Blok, Disney, Ekers, Freeman, Garcia, Gibson, Harnett, Henning,
  Jerjen, Kesteven, Knezek, Koribalski, Mader, Marquarding, Minchin, O'Brien,
  Oosterloo, Price, Putman, Ryder, Sadler, Stewart, Stootman, \&
  Wright}]{Meyer2004a}
Meyer, M.~J., {et~al.} 2004, \mnras, 350, 1195

\bibitem[{Mihos {et~al.}(2005)Mihos, Harding, Feldmeier, \&
  Morrison}]{Mihos2005a}
Mihos, J.~C., Harding, P., Feldmeier, J., \& Morrison, H. 2005, \apjl, 631, L41

\bibitem[{Miller {et~al.}(2002)Miller, Bregman, \& Knezek}]{Miller2002a}
Miller, E.~D., Bregman, J.~N., \& Knezek, P.~M. 2002, ApJ, 569, 134

\bibitem[{Morris {et~al.}(1993)Morris, Weymann, Dressler, Mccarthy, Smith,
  Terrile, Giovanelli, \& Irwin}]{Morris1993a}
Morris, S., Weymann, R., Dressler, A., Mccarthy, P., Smith, B., Terrile, R.,
  Giovanelli, R., \& Irwin, M. 1993, ApJ, 419, 524

\bibitem[{Morris {et~al.}(1991)Morris, Weymann, Savage, \&
  Gilliland}]{Morris1991a}
Morris, S., Weymann, R., Savage, B., \& Gilliland, R. 1991, ApJ, 377, L21

\bibitem[{{Norman} \& {Bryan}(1999)}]{Norman1999a}
{Norman}, M.~L., \& {Bryan}, G.~L. 1999, in Astrophysics and Space Science
  Library, Vol. 240, Numerical Astrophysics, ed. S.~M. {Miyama}, K.~{Tomisaka},
  \& T.~{Hanawa}, 19

\bibitem[{Oosterloo \& van~Gorkom(2005)}]{Oosterloo2005a}
Oosterloo, T., \& van~Gorkom, J. 2005, A{\&}A, 437, L19

\bibitem[{Ortiz~Gil {et~al.}(1999)Ortiz~Gil, Lanzetta, Webb, Barcons, \&
  Fernandez~Soto}]{Ortiz-Gil1999a}
Ortiz~Gil, A., Lanzetta, K.~M., Webb, J.~K., Barcons, X., \& Fernandez~Soto, A.
  1999, ApJ, 523, 72

\bibitem[O'Shea et al.(2004)]{O'Shea2004a} O'Shea, B.~W., Bryan, 
G., Bordner, J., et al.\ 2004, arXiv:astro-ph/0403044 

\bibitem[{{Paturel}(1979)}]{Paturel1979a}
{Paturel}, G. 1979, \aap, 71, 106

\bibitem[{Penton {et~al.}(2000)Penton, Stocke, \& Shull}]{Penton2000a}
Penton, S.~V., Stocke, J.~T., \& Shull, J.~M. 2000, \apjs, 130, 121

\bibitem[{Penton {et~al.}(2002)Penton, Stocke, \& Shull}]{Penton2002a}
---. 2002, \apj, 565, 720

\bibitem[{Penton {et~al.}(2004)Penton, Stocke, \& Shull}]{Penton2004a}
---. 2004, \apjs, 152, 29

\bibitem[{{Pierleoni} {et~al.}(2008){Pierleoni}, {Branchini}, \&
  {Viel}}]{Pierleoni2008a}
{Pierleoni}, M., {Branchini}, E., \& {Viel}, M. 2008, \mnras, 388, 282

\bibitem[{{Popping} \& {Braun}(2011)}]{Popping2011a}
{Popping}, A., \& {Braun}, R. 2011, \aap, 527, A90

\bibitem[{Prochaska {et~al.}(2011)Prochaska, Weiner, Chen, Mulchaey, \&
  Cooksey}]{Prochaska2011a}
Prochaska, J.~X., Weiner, B., Chen, H.-W., Mulchaey, J., \& Cooksey, K. 2011,
  ApJ, 740, 91

\bibitem[{Prochaska {et~al.}(2006)Prochaska, Weiner, Chen, \&
  Mulchaey}]{Prochaska2006a}
Prochaska, J.~X., Weiner, B.~J., Chen, H.-W., \& Mulchaey, J.~S. 2006, ApJ,
  643, 680

\bibitem[{Rauch(1998)}]{Rauch1998a}
Rauch, M. 1998, \araa, 36, 267

\bibitem[{Roncarelli {et~al.}(2006)Roncarelli, Ettori, Dolag, Moscardini,
  Borgani, \& Murante}]{Roncarelli2006a}
Roncarelli, M., Ettori, S., Dolag, K., Moscardini, L., Borgani, S., \& Murante,
  G. 2006, \mnras, 373, 1339

\bibitem[{Rosenberg {et~al.}(2003)Rosenberg, Ganguly, Giroux, \&
  Stocke}]{Rosenberg2003a}
Rosenberg, J.~L., Ganguly, R., Giroux, M.~L., \& Stocke, J.~T. 2003, \apj, 591,
  677

\bibitem[{Ryan-Weber(2006)}]{Ryan-Weber2006a}
Ryan-Weber, E.~V. 2006, \mnras, 367, 1251

\bibitem[{Salpeter \& Hoffman(1995)}]{Salpeter1995a}
Salpeter, E.~E., \& Hoffman, G.~L. 1995, \apj, 441, 51

\bibitem[{Schindler {et~al.}(1999)Schindler, Binggeli, \&
  B{\"o}hringer}]{Schindler1999a}
Schindler, S., Binggeli, B., \& B{\"o}hringer, H. 1999, \aap, 343, 420

\bibitem[{Sembach {et~al.}(2001)Sembach, Howk, Savage, Shull, \&
  Oegerle}]{Sembach2001a}
Sembach, K.~R., Howk, J.~C., Savage, B.~D., Shull, J.~M., \& Oegerle, W.~R.
  2001, ApJ, 561, 573

\bibitem[{{Shapley} \& {Ames}(1929)}]{Shapley1929a}
{Shapley}, H., \& {Ames}, A. 1929, Harvard College Observatory Bulletin, 865, 1

\bibitem[{Shull {et~al.}(2011)Shull, Smith, \& Danforth}]{Shull2011a}
Shull, J.~M., Smith, B.~D., \& Danforth, C.~W. 2011, arXiv:1112.2706

\bibitem[{Shull {et~al.}(2003)Shull, Tumlinson, \& Giroux}]{Shull2003a}
Shull, J.~M., Tumlinson, J., \& Giroux, M.~L. 2003, ApJ, 594, L107

\bibitem[{Smith {et~al.}(2000)Smith, Lucey, Hudson, Schlegel, \&
  Davies}]{Smith2000a}
Smith, R.~J., Lucey, J.~R., Hudson, M.~J., Schlegel, D.~J., \& Davies, R.~L.
  2000, \mnras, 313, 469

\bibitem[{Spitzer(1956)}]{Spitzer1956a}
Spitzer, J. 1956, \apj, 124, 20

\bibitem[{Thom \& Chen(2008)}]{Thom2008a}
Thom, C., \& Chen, H.-W. 2008, \apjs, 179, 37

\bibitem[{Thom {et~al.}(2011)Thom, Werk, Tumlinson, Prochaska, Meiring, Tripp,
  \& Sembach}]{Thom2011a}
Thom, C., Werk, J.~K., Tumlinson, J., Prochaska, J.~X., Meiring, J.~D., Tripp,
  T.~M., \& Sembach, K.~R. 2011, ApJ, 736, 1

\bibitem[{Tonnesen \& Bryan(2010)}]{Tonnesen2010a}
Tonnesen, S., \& Bryan, G.~L. 2010, ApJ, 709, 1203

\bibitem[{Tonnesen {et~al.}(2007)Tonnesen, Bryan, \& van
  Gorkom}]{Tonnesen2007a}
Tonnesen, S., Bryan, G.~L., \& van Gorkom, J.~H. 2007, ApJ, 671, 1434

\bibitem[{Tripp {et~al.}(2005)Tripp, Jenkins, Bowen, Prochaska, Aracil, \&
  Ganguly}]{Tripp2005a}
Tripp, T.~M., Jenkins, E.~B., Bowen, D.~V., Prochaska, J.~X., Aracil, B., \&
  Ganguly, R. 2005, ApJ, 619, 714

\bibitem[{Tripp {et~al.}(1998)Tripp, Lu, \& Savage}]{Tripp1998a}
Tripp, T.~M., Lu, L., \& Savage, B.~D. 1998, ApJ, 508, 200

\bibitem[{Tripp {et~al.}(2008)Tripp, Sembach, Bowen, Savage, Jenkins, Lehner,
  \& Richter}]{Tripp2008a}
Tripp, T.~M., Sembach, K.~R., Bowen, D.~V., Savage, B.~D., Jenkins, E.~B.,
  Lehner, N., \& Richter, P. 2008, \apjs, 177, 39

\bibitem[{Tripp {et~al.}(2002)Tripp, Jenkins, Williger, Heap, Bowers, Danks,
  Dav{\'e}, Green, Gull, Joseph, Kaiser, Lindler, Weymann, \&
  Woodgate}]{Tripp2002a}
Tripp, T.~M., {et~al.} 2002, ApJ, 575, 697

\bibitem[{Tully(1982)}]{Tully1982a}
Tully, R.~B. 1982, \apj, 257, 389

\bibitem[{Tully \& Shaya(1984)}]{Tully1984a}
Tully, R.~B., \& Shaya, E.~J. 1984, \apj, 281, 31

\bibitem[{Tumlinson {et~al.}(2011)Tumlinson, Werk, Thom, Meiring, Prochaska,
  Tripp, O'Meara, Okrochkov, \& Sembach}]{Tumlinson2011a}
Tumlinson, J., {et~al.} 2011, ApJ, 733, 111

\bibitem[{Urban {et~al.}(2011)Urban, Werner, Simionescu, Allen, \&
  B{\"o}hringer}]{Urban2011a}
Urban, O., Werner, N., Simionescu, A., Allen, S.~W., \& B{\"o}hringer, H. 2011,
  \mnras, 414, 2101

\bibitem[{van Gorkom {et~al.}(1996)van Gorkom, Carilli, Stocke, Perlman, \&
  Shull}]{Gorkom1996a}
van Gorkom, J.~H., Carilli, C.~L., Stocke, J.~T., Perlman, E.~S., \& Shull,
  J.~M. 1996, \aj, 112, 1397

\bibitem[{V{\'e}ron-Cetty \& V{\'e}ron(2006)}]{Veron-Cetty2006a}
V{\'e}ron-Cetty, M.-P., \& V{\'e}ron, P. 2006, A{\&}A, 455, 773

\bibitem[{White \& Rees(1978)}]{White1978a}
White, S.~D.~M., \& Rees, M.~J. 1978, \mnras, 183, 341

\bibitem[{Williger {et~al.}(2010)Williger, Carswell, Weymann, Jenkins, Sembach,
  Tripp, Dav{\'e}, Haberzettl, \& Heap}]{Williger2010a}
Williger, G.~M., {et~al.} 2010, \mnras, 405, 1736

\bibitem[{Wong {et~al.}(2006)Wong, Ryan-Weber, Garcia-Appadoo, Webster,
  Staveley-Smith, Zwaan, Meyer, Barnes, Kilborn, Bhathal, de~Blok, Disney,
  Doyle, Drinkwater, Ekers, Freeman, Gibson, Gurovich, Harnett, Henning,
  Jerjen, Kesteven, Knezek, Koribalski, Mader, Marquarding, Minchin, O'Brien,
  Putman, Ryder, Sadler, Stevens, Stewart, Stootman, \& Waugh}]{Wong2006a}
Wong, O.~I., {et~al.} 2006, \mnras, 371, 1855

\bibitem[{Yasuda {et~al.}(1997)Yasuda, Fukugita, \& Okamura}]{Yasuda1997a}
Yasuda, N., Fukugita, M., \& Okamura, S. 1997, \apjs, 108, 417

\bibitem[{York {et~al.}(2000)York, Adelman, Anderson, Anderson, Annis, Bahcall,
  Bakken, Barkhouser, Bastian, Berman, Boroski, Bracker, Briegel, Briggs,
  Brinkmann, Brunner, Burles, Carey, Carr, Castander, Chen, Colestock,
  Connolly, Crocker, Csabai, Czarapata, Davis, Doi, Dombeck, Eisenstein,
  Ellman, Elms, Evans, Fan, Federwitz, Fiscelli, Friedman, Frieman, Fukugita,
  Gillespie, Gunn, Gurbani, de~Haas, Haldeman, Harris, Hayes, Heckman,
  Hennessy, Hindsley, Holm, Holmgren, Huang, Hull, Husby, Ichikawa, Ichikawa,
  Ivezi{\'c}, Kent, Kim, Kinney, Klaene, Kleinman, Kleinman, Knapp, Korienek,
  Kron, Kunszt, Lamb, Lee, Leger, Limmongkol, Lindenmeyer, Long, Loomis,
  Loveday, Lucinio, Lupton, MacKinnon, Mannery, Mantsch, Margon, McGehee,
  McKay, Meiksin, Merelli, Monet, Munn, Narayanan, Nash, Neilsen, Neswold,
  Newberg, Nichol, Nicinski, Nonino, Okada, Okamura, Ostriker, Owen, Pauls,
  Peoples, Peterson, Petravick, Pier, Pope, Pordes, Prosapio, Rechenmacher,
  Quinn, Richards, Richmond, Rivetta, Rockosi, Ruthmansdorfer, Sandford,
  Schlegel, Schneider, Sekiguchi, Sergey, Shimasaku, Siegmund, Smee, Smith,
  Snedden, Stone, Stoughton, Strauss, Stubbs, SubbaRao, Szalay, Szapudi,
  Szokoly, Thakar, Tremonti, Tucker, Uomoto, Vanden~Berk, Vogeley, Waddell,
  Wang, Watanabe, Weinberg, Yanny, \& Yasuda}]{York2000a}
York, D.~G., {et~al.} 2000, \aj, 120, 1579

\end{thebibliography}

\end{document}